\documentclass[12pt,oneline,a4paper]{ouparticle}

\usepackage{dsfont}
\usepackage{color,soul}
\usepackage[normalem]{ulem}
\usepackage[flushleft]{threeparttable} 
\usepackage{icomma}
\usepackage{makecell}
\usepackage[flushleft]{threeparttable}
\usepackage{multirow}
\usepackage{listings}
\usepackage{apacite}
\usepackage{latexsym}
\usepackage{amsmath}
\usepackage[authoryear]{natbib}
\usepackage{booktabs}
\usepackage{amsfonts}
\usepackage{caption}
\usepackage{appendix}
\usepackage{amssymb}
\usepackage{bm}
\usepackage{xr}
\usepackage{rotating}
\DeclareMathOperator*{\argmin}{arg\,min}
\usepackage{psfrag}
\usepackage{algorithm,algorithmic}
\usepackage{graphicx}
  \algsetup{
    linenodelimiter={.}
  }
\usepackage{mathtools}
\usepackage{float}
\usepackage{xcolor}
\usepackage{soul}
\usepackage{subcaption}

\DeclarePairedDelimiter\floor{\lfloor}{\rfloor}

\begin{document}

\title{Smoothed pseudo-population bootstrap methods\\ with applications to finite population quantiles}

\author{%
\name{Vanessa McNealis}
\address{University of Glasgow, School of Mathematics and Statistics, Glasgow, Scotland,\\
United Kingdom, G12 8QW}
\email{vanessa.mcnealis@glasgow.ac.uk}
\and
\name{Christian Léger}
\address{Université de Montréal, Département de mathématiques et de statistique, Montréal, Québec\\
Canada, H3T 1N8}
\email{christian.leger@umontreal.ca}}

\abstract{This paper introduces smoothed pseudo-population bootstrap methods for the purposes of {mean squared error} estimation and for constructing confidence intervals for finite population quantiles. In an i.i.d.~context, it has been shown that resampling from a smoothed estimate of the distribution function instead of the usual empirical distribution function can improve the convergence rate of the bootstrap {mean squared error} estimator of a sample quantile. We extend the smoothed bootstrap to the survey sampling framework by implementing it in pseudo-population bootstrap methods for high entropy, single-stage survey designs, such as simple random sampling without replacement, Poisson sampling, and {randomized systematic proportional-to-size sampling}. Given a kernel function and a bandwidth, it consists of smoothing the pseudo-population from which bootstrap samples are drawn using the original sampling design. Given that the implementation of the proposed algorithms requires the specification of the bandwidth, we develop a plug-in selection method along with a grid search selection method based on a bootstrap estimate of the mean squared error. Simulation results suggest that the smoothed approach offers improved efficiency compared to the standard pseudo-population bootstrap for estimating the {uncertainty} of a quantile estimator together with mixed results regarding confidence interval coverage.}

\date{\today}


\maketitle

\subsubsection*{Statement of significance}

In survey sampling, {uncertainty} estimation for finite population quantiles can be challenging, as standard bootstrap methods have shown to be inefficient. To address this, this manuscript presents a novel bootstrap methodology for uncertainty quantification for the survey sampling context inspired by the smoothed bootstrap introduced in classical statistics. The proposed methodology is particularly valuable for national statistical agencies and researchers working with skewed data, where quantile estimation is essential. By introducing a smoothed pseudo-population bootstrap algorithm, the paper addresses a key limitation in {mean squared error} estimation for finite population quantiles, offering improved efficiency compared to standard methods. Additionally, the development of data-driven bandwidth selection methods enhances the practical application of the technique. The findings have broad implications for the statistical community, improving the accuracy of {uncertainty} estimation and confidence interval construction in complex survey designs.

\section{Introduction}
\label{sec1}

The sample median is a robust and well-defined measure of central tendency, known for being consistent for the population median. Quantiles are generally appropriate descriptors of heavily skewed distributions such as population income distributions. A well known fact is that quantiles are \textit{not} smooth statistics, in the sense that they are not differentiable functions of means and their variance depends on local properties of the underlying distribution of the observations \citep{efron1993introduction}, which complicates variance estimation for these statistics. In the context of survey sampling, while analytical expressions for the asymptotic variance of a sample quantile are available for some designs \citep{francisco1991quantile}, in practice, national statistical agencies routinely use bootstrap methods to compute finite sample {uncertainty} estimators and construct confidence intervals for population quantiles. 


With survey data, bootstrap methods must be designed so as to account for the effect of the sampling design on the variability of estimators. \cite{mashreghi2016survey} classified existing bootstrap methods for survey data into three groups: pseudo-population, direct, and bootstrap weights. In pseudo-population bootstrap methods, a pseudo-population is first constructed by repeating units in the sample and bootstrap samples are drawn from the pseudo-population using the original sampling design \citep{gross1980median, booth1994bootstrap, chauvet2007bootstrap}. By emulating the initial sampling design to draw bootstrap samples, pseudo-population schemes produce variance estimators which naturally capture finite population correction factors \citep{mashreghi2016survey}. This attractive property has led many researchers and statistical agencies to actively pursue this research area \citep{sitter1992comparing, chao1994maximum, saigo2010comparing, chen2019pseudo, chen2022comparison}. However, many of the simulations for quantiles such as the median have shown relatively poor results in small to moderate samples whether it be for {mean squared error} estimation or for confidence intervals; see, for instance, \cite{sitter1992comparing} for stratified simple random sampling or \cite{saigo2010comparing} for three-stage stratified simple random sampling. This is attributable to the lack of smoothness in these statistics, which manifests in poor support of resulting bootstrap distributions.

One approach that sometimes helps in classical statistics is smoothing. The smoothed bootstrap was introduced by \cite{efron1979} upon examining the problem of estimating the variance of the sample median for independent and identically distributed (i.i.d.) observations. It consists of adding a small amount of random noise to each bootstrap sample so as to enrich the support of the bootstrap distribution. It may be shown that the usual (unsmoothed) bootstrap estimate of the {mean squared error} of the sample median has a relative error of order $n^{-1/4}$ as sample size $n$ increases \citep{hall1988exact}. Provided the use of a second-order, nonnegative kernel, \cite{hall1989smoothing} showed that the order of the relative error of the bootstrap estimate of the {mean squared error} of the sample quantile can be improved from $n^{-1/4}$ to $n^{-2/5}$ by smoothing the empirical distribution function from which bootstrap samples are taken. In contrast, for linear functionals, the smoothed bootstrap may only have a second order improvement in a mean squared error sense \citep{silverman1987bootstrap}. To our knowledge, the idea of smoothing the bootstrap has not been exploited in survey sampling even though it is clearly feasible within the pseudo-population framework.

In this work, we address this important gap by putting forward a smoothed pseudo-population bootstrap methodology for single-stage, high entropy designs, such as {simple random sampling without replacement (SRSWOR), Poisson sampling, and randomized systematic proportional-to-size (PPS) sampling}. This methodology requires the specification of a smoothing parameter or bandwidth, which controls the degree of smoothing in the pseudo-population and, in turn, in the bootstrap samples. Throughout the paper, we focus on the problem of {mean squared error} estimation and confidence interval construction for finite population quantiles. Since the optimal bandwidth is typically unknown, a large part of our discussion centers around data-driven selection approaches for the bandwidth. Our first proposal is a double bootstrap selection procedure, which selects the optimal bandwidth in a grid search based on a bootstrap estimate of the quadratic risk for a bootstrap mean squared error estimator. For the case of SRSWOR, we introduce a plug-in method which relies upon the expression of the optimal bandwidth for the variance of the sample quantile derived in the i.i.d. setting \citep{hall1989smoothing}. 

The paper is structured as follows. In Section \ref{sec2}, we introduce the survey sampling framework and relevant notation. In Section \ref{sec3}, we describe our proposed smoothed pseudo-population algorithm for high entropy survey designs assuming a fixed, known bandwidth value. In Section \ref{sec4}, we describe two data-driven bandwidth selection approaches, a double bootstrap procedure and a plug-in approach. In Section \ref{sec5}, we provide the methodology setup of our simulation studies and empirically compare the smoothed pseudo-population bootstrap procedure based on the two bandwidth selection approaches to the standard, unsmoothed procedure. We close with a general discussion in Section \ref{sec6}. {Our discussions focus on single-stage sampling designs but the methodological development can be used for more general sampling designs.}

\section{Notation and Survey Sampling Framework}
\label{sec2}

Let $U$ be a finite population of $N$ units, labeled by integers $1, \ldots, N$, and $y$ a study variable. For each unit $i$, denote by $y_i$ the corresponding value of $y$. Let $\theta = \theta(U)$ denote a finite population parameter.

To estimate $\theta$, a sample $S \subset U$ of size $n_S$ is drawn according to a sampling design $p(S)$ with sampling fraction $f = n_S/N$, where $n_S$ may be a random variable. We denote the first-order inclusion probabilities as $\pi_i = P(i \in S)$. Let $\mathbb{E}_p$ and ${V}_p$ denote the expectation and the variance with respect to the sampling design $p(\cdot)$. The expected sample size for design $p(\cdot)$ is therefore $n = \mathbb{E}_p[n_s]$. 

While the smoothed pseudo-population bootstrap methodology that we will introduce could be applied to the estimation of different finite population parameters, we focus on one which has the potential to benefit the most. Consider  $\theta = \xi_p, p \in (0,1)$, namely the $p$-th level finite population quantile given by 
\begin{equation}
\label{eqn:finitepopquant}
{\xi}_p = \begin{cases} \ y_{(k)}, & F_N( y_{(k-1)}) <p <  F_N(y_{(k)}), \\
\ \frac{1}{2}(y_{(k)} + y_{(k+1)} ), &  F_N(y_{(k)}) = p, \end{cases}
\end{equation}
where $y_{(1)}\le y_{(2)}\le \ldots\le y_{(N)}$ correspond to the ordered values in the population and $F_N(t) = N^{-1} \sum_{i=1}^N \mathds{1}(y_i \leq t)$ denotes the finite population distribution function.
A design-based estimator $\hat{\xi}_p$ of $\xi_p$ is obtained by replacing $F_N$ in (\ref{eqn:finitepopquant}) by an estimator of the distribution function commonly attributed to \cite{hajek1971comment} defined by
\begin{equation}
\label{eqn:estimatorquant}
\hat{F}(t) = \left( \sum_{i \in s} \pi_i^{-1}\right)^{-1}\sum_{i \in s} \pi_i^{-1}\mathds{1}(y_i \leq t). \nonumber
\end{equation}
Note that under the simple random sampling without replacement design (SRSWOR), in which $\pi_i = n/N \ \forall i \in U$, the design-based quantile estimator $\hat{\xi}_p$ and the usual $p$-th level sample quantile coincide. 

\section{Smoothed Pseudo-Population Bootstrap Methods}
\label{sec3}

In this section, we propose a smoothed version of an existing class of design-based bootstrap methods --- pseudo-population bootstrap methods --- that may be used to estimate the distribution of estimators of finite population parameters. We first revisit the smoothed bootstrap under the i.i.d.~model before moving on to the adaptation of this method to the survey sampling setting.

\subsection{The Smoothed Bootstrap in a Classical Setting}
Let $Y_1, Y_2, \ldots, Y_n$ denote a random sample of $n$ i.i.d.~observations drawn from an unknown probability distribution $F_0$. Also, let the statistic $\hat{\theta} \equiv \hat{\theta}(Y_1, Y_2, \ldots, Y_n)$ denote a sample estimate of a parameter of interest ${\theta} = {\theta}(F_0)$, such as the mean or the median of the distribution. Note that, although the same notation is used across the i.i.d.~and the survey sampling settings for simplicity, these quantities should not be confused with the aforementioned survey estimator and finite population parameter, respectively. Now, consider the problem of estimating an attribute of the sampling distribution function of $\hat{\theta}$ centered at $\theta$, denoted $J_n(t, F_0) = \mathrm{Prob}_{F_0}(\hat{\theta}-{\theta} \leq t)$, such as the variance or a quantile of this distribution. 
 
The bootstrap method consists of substituting an estimate of $F_0$, denoted $\hat{F}$, into the functional $J_n(t, \cdot)$ in a way to obtain $J_n(t,\hat{F})$. 
The nonparametric bootstrap \citep{efron1979} plugs in the empirical distribution function $\hat{F}_n$ which attributes a weight $1/n$ to each value $Y_i$. More often than not, the estimated sampling distribution $J_n(t,\hat{F}_n)$ is itself approximated through a resampling algorithm. In doing so, a finite number of bootstrap samples of the form $Y^*_1, Y^*_2, \ldots, Y^*_n$ are drawn successively with replacement from the initial sample and the approximation of the bootstrap distribution is formed by the resulting collection of bootstrap statistics, each given by $\hat{\theta}^* = \hat{\theta}(Y^*_1, Y^*_2, \ldots, Y^*_n)$. 

In some instances, attributing a certain amount of smoothness to $F_0$ can be beneficial. The smoothed bootstrap (Efron, 1979) consists of plugging a smoothed version $\hat{F}$ of $\hat{F}_n$ into the functional  $J_n(t, \cdot)$. One possibility is to let $\hat{F}_h$ be the Parzen-Rosenblatt estimate of the cumulative distribution function $F_0$ \citep{parzen1962estimation}, defined as follows
\begin{equation}
\label{eqn:kernel}
\hat{F}_h(t) = n^{-1}\sum_{i=1}^n K\left\{(Y_i - t)/h\right\}, 
\end{equation}
where $K(x) = \int_{-\infty}^x k(t)dt$, $k$ is a kernel function and $h > 0$ is a smoothing parameter or bandwidth that controls the degree of smoothing. Furthermore, we restrict $k$ to be a probability density function satisfying  $\int tk(t)dt = 0$ and $\int t^2k(t)dt  < \infty$. Given the constraints on $k$, the bootstrap distribution $J_n(t,\hat{F}_h)$ can be approximated through a resampling algorithm. For implementation purposes, we may utilize the fact that the kernel estimator $\hat{F}_h(t)$ in (\ref{eqn:kernel}) is the convolution of the empirical distribution function $\hat{F}_n$ with the cumulative distribution function $K$ with smoothing parameter $h$. With this in mind, the task of drawing a bootstrap sample from $\hat{F}_h$ becomes rather simple: with $Y^*_i$ as defined above, let
 $$X^*_i = Y^*_i + h\varepsilon_i^*,$$
 where $\varepsilon_i^* \sim K$, then $X^*_i \sim \hat{F}_h$, $i=1,\ldots,n$.

\subsection{An Extension to the Survey Sampling Framework}
\label{section:smoothedbootstrapsurvey}
We propose an extension of the smoothed bootstrap procedure to the finite population setting through the class of pseudo-population bootstrap methods. Such methods consist of creating a pseudo-population from the units in the sample and taking bootstrap samples according to the same sampling design that led to the original sample. In that regard, the overarching principle behind pseudo-population bootstrap methods is analogous to that of the nonparametric bootstrap in the sense that it can be equally described as a \textit{plug-in rule}. However, in the finite population setting, the unknown quantity is the population $U$ instead of a distribution $F_0$. Therefore, the sampling distribution of an estimator $\hat{\theta}$ for a fixed sample size design is henceforth denoted by $J_{n}(t, U) = \mathrm{Prob}_{U}(\hat{\theta}-\theta \leq t)$. The substitution of $U$ by a pseudo-population $U^*$, a sample estimate of $U$, into the functional $J_{n}(t, \cdot)$ leads to the bootstrap distribution $J_{n}(t, U^*)$.

The construction of $U^*$ typically depends on the sampling design, which determines the first-order inclusion probabilities of the units in $S$ and thus their relative importance within the pseudo-population. \cite{booth1994bootstrap} introduced a method for simple random sampling without replacement (SRSWOR), which can easily be extended to stratified simple random sampling. An algorithm for Poisson sampling was first described by \cite{chauvet2007bootstrap}. Both of these designs belong to the class of high entropy designs and the algorithm formulated by \cite{chauvet2007bootstrap} may encompass other designs in this class, {such as the Rao-Sampford method \citep{rao1965two, sampford1967sampling} or the randomized systematic PPS design \citep{hartley1962sampling}}.

 A general smoothed pseudo-population resampling algorithm for high entropy designs is described in Algorithm \ref{algo:smoothedppb}, which results into a smoothed bootstrap estimate of the sampling distribution, $J_{n}(t, U^*_h)$. In the proposed method, the pseudo-population is smoothed prior to resampling and the resulting smoothed pseudo-population $U^*_h$ corresponds to the convolution of the observations in $U^*$ and a random variable with cumulative distribution function $K_h$, where $K_h(t) = K(t/h)$, $t \in \mathbb{R}$. 
If no random noise is added to the observations in the pseudo-population, then Algorithm \ref{algo:smoothedppb} coincides with the (unsmoothed) pseudo-population algorithm for unequal single-stage probability designs (UEQPS PPB) explicited by \cite{mashreghi2016survey}. The main difference between Algorithm \ref{algo:smoothedppb} and the UEQPS PPB algorithm lies in Step 3, in which the smoothing of $U^*$ occurs. Moreover, while steps 4 to 7 also appear in the standard method, the bootstrap estimates in the smoothed method are now indexed by $h$.

\begin{algorithm}[h]

    \caption{Smoothed Pseudo-Population Bootstrap for UEQPS Designs}
  \begin{algorithmic}[1]
   \vspace*{0.5cm}
    \STATE Form $U^f$, the fixed part of the pseudo-population, by replicating each pair $(y_i, \pi_i)$ a total of $\floor{\pi_i^{-1}}$ times, with $\floor{x}$ being the largest integer less or equal to $x$.
    \STATE Complete the pseudo-population by drawing $U^{*c}$ according to the original survey design with inclusion probability equal to $\pi_i^{-1} - \floor{\pi_i^{-1}}$ for unit $(y_i, \pi_i), \ i \in S$, leading to the pseudo-population $U^* = U^f \cup U^{c*} = \{(y^*_i, \pi^*_i)\}_{ i = 1, \ldots, N^* }$ with possibly random size $N^*$, {where $(y^*_i, \pi^*_i)$ corresponds to one of the original pairs of values of the variable and first-order inclusion probability in $S$.}
    
    \STATE To obtain a smoothed pseudo-population $U^*_h$, compute $y^*_{i,h} = y^*_i + h\varepsilon^*_i$, where $\varepsilon^*_i \stackrel{\text{i.i.d.}}{\sim} K$, $i = 1, \ldots, N^*$, and $h$ is the smoothing parameter.
   \STATE Compute the smoothed bootstrap parameter, $\theta^*_h = \theta(U^*_h)$, on the pseudo-population $U^*_h$. 
      \STATE {Using the original sampling design, generate a bootstrap sample $S^*_h $ from $U^*_h$, but with inclusion probability $\pi'_i$ for unit $i \in U^*_h, i = 1, \ldots, N^*$, as defined in the sequel.}
      \STATE Compute the smoothed bootstrap estimator, given by $\hat{\theta}^*_h = \theta(S^*_h)$.
      \STATE {For $b=1,\ldots, B$, with $B$ large enough, repeat the steps above so as to obtain the following distributions of bootstrap parameters and estimates:
\[ (\theta_{1,h}^*, \ldots, \theta_{B,h}^*)' \quad \text{and} \quad (\hat{\theta}_{1,h}^*, \ldots, \hat{\theta}_{B,h}^*)'.\]}
  \end{algorithmic}
  \label{algo:smoothedppb}
\end{algorithm}

The pair $(y_i, \pi_i)$ in Algorithm \ref{algo:smoothedppb} denotes the $i$-th sampled measurement, $i \in S$, along with its corresponding first-order inclusion probability. A special case of Algorithm \ref{algo:smoothedppb} is the method for SRSWOR, in which $\pi_i \equiv  n/N$. It follows that in Step 1, the fixed part of the pseudo-population $U^f$ is formed by replicating each sample unit $k = \floor{N/n}$ times and a SRSWOR of size $N - nk$ is drawn from $S$ in Step 2 to form $U^{*c}$ \citep{booth1994bootstrap}. In contrast to the method for SRSWOR, in which the pseudo-population is of fixed size $N$, the method for Poisson sampling has a random pseudo-population size. To complete the pseudo-population in Step 2, each unit in $S$ is included independently in $U^{*c}$ with probability $\pi_i^{-1} - \floor{\pi_i^{-1}}$, where $\pi_i$ is the first-order inclusion probability of the original Poisson sampling scheme \citep{chauvet2007bootstrap}. {In the case of SRSWOR or Poisson sampling, we set $\pi_i' = \pi^*_i \ \forall i \in U^*_h$ to perform bootstrap sampling in Step 5, that is, a unit is included in $S^*$ according to the survey design with first-order inclusion probability corresponding to the original inclusion probability in the population.}

{In the case of PPS designs, such as randomized systematic PPS sampling, $\pi'_i$ may differ from $\pi_i^*$. A PPS design often entails defining first-order inclusion probabilities as a function of a size variable $x$, which is assumed to be available for the whole population and to be correlated with the variable of interest $y$. 
Generally, $\pi_i = n p_i$, where $p_i = x_i/t_x$ with $t_x = \sum_{i \in U} x_i$, $i \in U$, and $n$ is the target sample size, such that $\sum_{i \in U} p_i = 1$. Since the size distribution in $U^*_h$ is not the same as in the original population, the first inclusion probabilities used in Step 5 are modified to $\pi_i' = n \pi_i^*/\sum_{i \in U^*_h} \pi_i^*$, where $\pi_i^* = n p_i^*$. It is generally assumed that $p_i \leq 1/n $ or, equivalently, $\pi_i \leq 1$, $\forall \ i \in U$. Otherwise, if some values $x_i$ are too large leading to $\pi_i > 1$ for some $i \in U$, corresponding inclusion probabilities are rounded to 1 and first-order inclusion probabilities are recalculated as
$$ \pi_i = \min\left(1, h^{-1}(n) p_i  \right), \quad i \in U,$$
where $h(z) = \sum_{i \in U} \min \left( z p_i, 1\right)$ \citep{deville1998unequal, chauvet2007bootstrap}. This procedure also applies to the inclusion probabilities to perform the bootstrap sampling in Step 5 in the case of a violation of the requirement $\pi_i' \leq 1 \ \forall i \in U^*_h$. }


{Finally, as noted by \cite{mashreghi2016survey} for the unsmoothed case, the smoothed UEQPS PPB algorithm can easily be extended to stratified simple random sampling without replacement by applying a resampling method independently within strata, where the bandwidth $h$ could vary by stratum depending on stratum size. If the target of inference is a finite population quantile, the smoothing parameter could for instance be of form $h=Cn_l^{-1/5}$ for stratum $l$, $l=1, \ldots, L$, with $n_l$ being the sample size for stratum $l$ and $L$ is the number of strata (see Section \ref{sec4} for a discussion of the choice of smoothing parameter). This way, a number of $L$ smoothed pseudo-populations would be formed by replicating the units in each stratum as described in Algorithm 1. A smoothed bootstrap sample $S^*_h$ would be taken as the union of SRSWORs drawn independently from the $L$ smoothed pseudo-populations.  In the particular case of the $p$-th level quantile, the smoothed bootstrap quantile estimator $\hat{\xi}^*_{p,h}$ would be computed by inverting the following smoothed estimator of the cumulative distribution function}

$$ {\hat{F}^*_h(t) = \frac{1}{N} \sum_{l=1}^L \sum_{i \in S^*_{h,l}} \frac{N_l}{n_l} \mathds{1}(y^*_{i,h,l} \leq t),}$$ {where $y^*_{i,h,l}$ is the $i$-th observation of the study variable in the smoothed bootstrap sample drawn from the $l$-th stratum of the smoothed pseudo-population, which are respectively denoted $S^*_{h,l}$ and $U^*_{h,l}$. Likewise, the smoothed bootstrap parameter ${\xi}^*_{p,h}$ would be obtained by inverting $F_h^*(t)$, where ${F}^*_h(t) = N^{-1} \sum_{l=1}^L \sum_{i \in U^*_{h,l}} \mathds{1}(y^*_{i,h,l} \leq t).$ By replicating the sampling process $B$ times, we would obtain the smoothed bootstrap distributions of quantile estimators $(\hat{\xi}^*_{1,h}, \ldots, \hat{\xi}^*_{B,h})$ and parameters $({\xi}^*_{1,h}, \ldots, {\xi}^*_{B,h})$.}

\subsection{Design-Based Mean Squared Error and Confidence Interval Estimation}
\label{subsection:veic}
Assuming that $\hat{\theta}^*_h$ is design-unbiased for $\theta^*_h$, a smoothed bootstrap variance estimator of $\mathrm{Var}_p(\hat{\theta})$ is given by
\begin{align}
\label{eqn:bootstrap_estimator}
  V^*(\hat{\theta}^*_h) = \mathbb{E}_{u^*}\left[V^*_{p^*}(\hat{\theta}^*_h\ \rvert U^*_h)\right],
\end{align}
where the subscripts $u^*$ and $p^*$ denote the random processes of pseudo-population $U^*$ completion and resampling respectively. {The estimator $V^*(\hat{\theta}^*_h)$ is the average over different pseudo-populations of the sampling variability of the bootstrap estimator $\hat{\theta}^*_h$.} A Monte Carlo approximation of (\ref{eqn:bootstrap_estimator}) is obtained by exploiting the bootstrap distributions given in Step 7 of Algorithm 1:
\begin{align}
    \label{eqn:bootstrap_approximation}
    \hat{V}_h = \frac{1}{B}\sum_{b=1}^B \left(\hat{\theta}^*_{b,h} - \theta^*_{b,h} \right)^2.
    \end{align}
 Should $\hat{\theta}^*_h$ be biased as an estimator of $\theta^*_h$, the estimator in (\ref{eqn:bootstrap_approximation}) would rather be an approximation of the smoothed bootstrap estimate of the mean squared error of $\hat{\theta}$. {Note that this is the case of the quantile estimator in the context of a finite population, where the bias is of order $1/n$ (see \cite{francisco1991quantile} for the case of SRSWOR). Therefore, in simulation studies for finite population quantiles in Section \ref{sec5}, the accuracy of $\hat{V}_h$ is assessed with respect to the mean squared error as opposed to the variance. Section \ref{appendix:smoothedppbvariance} of the Appendix describes a different more computationally intensive smoothed pseudo-population bootstrap algorithm tailored for variance estimation. Please note that we will continue to use the notation $\hat V_h$ even though it is an estimator of the mean squared error instead of the variance. }

 We consider two approaches to construct $1-\alpha$ level bootstrap confidence intervals for $\theta$. The first method consists of computing the mean squared error estimate of $\hat{\theta}$ as in (\ref{eqn:bootstrap_approximation}) and of using the normal approximation of the distribution of $\hat{\theta}$ in a way to obtain the $1-\alpha$ level asymptotic bootstrap confidence interval 
 \begin{equation}
 \label{eqn:IC_normal}
\left[\hat{\theta} - z_{1-\frac{\alpha}{2}}\sqrt{\hat{V}^*_h},\hat{\theta} + z_{1-\frac{\alpha}{2}}\sqrt{\hat{V}^*_h} \right],
\end{equation}
where $z_{\beta}$ is the $\beta$-quantile of the standard normal distribution. 

The second approach, a basic bootstrap confidence interval \citep{davison1997bootstrap}, makes direct use of the bootstrap distributions in step 7 of Algorithm \ref{algo:smoothedppb}. Recall that, in the finite population setting, we denote the sampling distribution of $\hat{\theta}$ as $J_n(t, U) = \mathrm{Prob}_U\left(\hat{\theta} - \theta \le t \right)$. The following probabilistic statement dictates the form of the basic interval
$$P\left( \hat{\theta} - J_n^{-1}\left(1-\alpha/2, U \right) \leq \theta \leq \hat{\theta} -  J_n^{-1}\left(\alpha/2, U \right)\right) = \alpha,
$$ where the quantiles $J_n^{-1}\left(1-\alpha/2, U \right)$ and $J_n^{-1}\left(\alpha/2, U \right)$ are unknown. By substituting unknown quantities with their bootstrap analogues, the smoothed $1-\alpha$ level basic bootstrap interval is given by
\begin{equation}
\label{eqn:IC_base}
\left[\hat{\theta} - J_n^{-1}\left(1-\alpha/2, U^*_h \right),\  \hat{\theta} - J_n^{-1}\left(\alpha/2, U^*_h \right) \right],
\end{equation}
where $J_n \left(t, U^*_h \right) = \mathrm{Prob}_{U^*_h}(\hat{\theta}^*_h - \theta^*_h \ \leq t)$. It is worth mentioning that the value of the bootstrap parameter $\theta^*_h$ will change across different boostrap pseudo-populations.

\section{Choice of Bandwidth}
\label{sec4}
The question of how much to smooth is invariably a delicate one in any smoothing problem. Special care must be given to the choice of the bandwidth $h$ in the smoothed pseudo-population bootstrap methods, in which a suboptimal value may not only offer no or little improvement, but even have deleterious effects on bootstrap estimates compared with the standard algorithm.

While we know that the bandwidth must converge to 0 as the sample size increases, the optimal rate of convergence is unclear at first glance. In the i.i.d.~context, the asymptotic variance of the $p$-th quantile is $p(1-p)/f^2_0(\xi_p)$. \cite{hall1989smoothing} showed that
the smoothed bootstrap estimate of $\mathrm{Var}(\hat{\xi}_p)$ converges to the asymptotic variance at a faster rate than if unsmoothed, with the optimal rate being attained by a bandwidth of order $n^{-1/5}$. Consider now the survey sampling context. If we assume a superpopulation model, i.e., that the finite population is the result of $N$ i.i.d.~draws from distribution $F_0$ with density $f_0$, {\cite{chatterjee2011asymptotic} showed that for SRSWOR, the asymptotic variance of the $p$-th quantile is $(1-f)p(1-p)/f_0^2({\xi}_p)$ where ${\xi}_p$ is the $p$-th quantile of $F_0$ and $f$ is the sampling fraction. This suggests that, in the survey sampling setting at the least for SRSWOR, we should also consider smoothing parameters of the form $h=Cn^{-1/5}$ for a certain value $C>0$.

We consider two data-driven approaches for the empirical choice of $h$ in the implementation of Algorithm \ref{algo:smoothedppb}. The first one consists of computing a bootstrap estimate of the mean squared error of the bootstrap {mean squared error} estimator of the $p$-th quantile on a grid of values of constants $\mathcal{C}$ (or equivalently a grid of bandwidths $\mathcal{H}$). This results in a double bootstrap. 

Alternatively, in the i.i.d.~case, \cite{hall1989smoothing} have derived the optimal constant for quantile variance estimation, which depends on the true distribution. {Given the result of \cite{chatterjee2011asymptotic} for SRSWOR under the superpopulation model, assuming that the optimal constant is the same, and using sample estimates of the unknown quantities, we also introduce a plug-in bandwidth estimator. The latter methodology does not easily generalize to other sampling plans.}

{While the plug-in bandwidth selection approach assumes a correct order of $n^{-1/5}$ for the bandwidth, as it targets the optimal constant derived by \cite{hall1989smoothing}, the proposed bootstrap bandwidth selection method can be carried out without knowledge of the correct order for $h$. As we explain later in this section, an arbitrary grid $\mathcal{H}$ can be initialized and adjusted until a global minimum of the estimated mean squared of $\hat{V}^*_h$ is found. Indeed, while using a particular rate such as $n^{-1/5}$ as the scale for identifying the bandwidth can be useful, in a given problem the user must still find a grid $\mathcal{C}$ leading to a grid $\mathcal{H}$ that will contain a global minimum. Hence identifying the correct asymptotic rate is not crucial in practice.}

\subsection{Double Bootstrap Procedure}

The main suggested method is based on the use of the bootstrap to obtain an empirical estimate of risk, a principle that was studied in broad terms by \cite{leger1990bootstrap} and by \cite{hall1990using}. \cite{de1992smoothing} applied this principle to provide a general data-driven bandwidth selection method labeled as the double bootstrap for the smoothed bootstrap within the i.i.d.~model. Simulation results in the classical setting for the estimation of the sample median variance support the efficacy of the smoothed bootstrap with bandwidth chosen by the double bootstrap over the unsmoothed bootstrap \citep{de1992smoothing}.

Although this method is a priori applicable to other functionals, we describe a double bootstrap approach specifically designed to optimize the stability of the {mean squared error} estimator $V^*(\hat{\theta}^*_h)$ in (\ref{eqn:bootstrap_estimator}) of an arbitrary survey estimator $\hat{\theta}$. Consequently, consider the risk function
\begin{equation}
\label{eqn:risk}
{\mathrm{MSE}}\left(V^*(\hat{\theta}^*_h) \right) = \mathbb{E}_p\left[\left\{V^*(\hat{\theta}^*_h) - V_p(\hat{\theta})\right\}^2 \right],
\end{equation}
namely the mean squared error of the bootstrap estimate $V^*(\hat{\theta}^*_h)$ under sampling design $p(\cdot)$. A bootstrap estimate of (\ref{eqn:risk}) is given by
\begin{equation}
\label{eqn:risk_bootstrap}
\widehat{\mathrm{MSE}}\left(V^*(\hat{\theta}^*_h)\right) = \mathbb{E}^*\left[\left\{V^{**}(\hat{\theta}^{**}_h) - V^*(\hat{\theta}^*_g) \right\}^2 \right],
\end{equation}
where $ g \geq 0$ and $V^{**}(\hat{\theta}^{**}_h)=\mathbb{E}_{u^{**}}\left[V^{**}_{p^{**}}(\hat{\theta}^{**}_h\ \rvert U^{**}_h)\right]$ denotes the {(double)} bootstrap estimator of $V^*(\hat{\theta}^*_h)$. While the latter estimator corresponds to the {mean squared error} of the bootstrap sampling distribution $J_{n}(t, U^*_h) = \mathrm{Prob}_{U^*_h}(\hat{\theta}^*_h-\theta^*_h \leq t)$, the former is the {mean squared error} of the (double) bootstrap distribution $J_{n}(t, U^{**}_h) = \mathrm{Prob}_{U^{**}_h}(\hat{\theta}^{**}_h-\theta^{**}_h \leq t)$, where $U^{**}_h$ is the smoothed pseudo-population constructed from a random sample taken from $U^*_g$ and $\theta^{**}_h$ and $\hat{\theta}^{**}_h$ are the corresponding bootstrap parameter and estimator, respectively. A natural choice for $g$ is simply $g = h$. The value $\hat{h}$ that minimises the expression in (\ref{eqn:risk_bootstrap}) may then be chosen as the smoothing parameter. 

To fix ideas, we illustrate the double bootstrap simulation process for selecting a bandwidth that minimizes a finite approximation of the empirical risk (\ref{eqn:risk_bootstrap}) with $g = h$ among a suitable grid of bandwidths $\mathcal{H}=\{ h_1, h_2, \ldots h_m\}$, where $h_i = C_in^{-1/5},\ i =1,\ldots,m$. It is done by generating a bootstrap distribution of the estimator $\hat{V}^*_h$ in (\ref{eqn:bootstrap_approximation}) from the $B$ first-level bootstrap samples in Algorithm \ref{algo:smoothedppb} for each value $h \in \mathcal{H}$. Recall that in step 6 of Algorithm \ref{algo:smoothedppb}, a smoothed bootstrap sample $S^*_h$ is drawn from $U^*_{h}$. Thus, to obtain the desired second-level bootstrap distribution, we apply Algorithm \ref{algo:smoothedppb} to $S^*_h$ in a nested fashion with bandwidth $h$ and $D$ bootstrap replicates, leading to the collections $(\theta^{**}_{1,h}, \ldots, \theta^{**}_{D,h} )'$ et $(\hat{\theta}^{**}_{1,h}, \ldots, \hat{\theta}^{**}_{D,h} )'$. By repeating this two-stage process $B$ times, we may obtain $\hat{V}^*_h$ as well as the collection of second-level bootstrap variance estimates $(\hat{V}^{**}_{1,h}, \ldots, \hat{V}^{**}_{B,h})'$ for each value $h \in \mathcal{H}$, where \begin{equation}
\label{eqn:doublevar}
\hat{V}^{**}_{b,h} = \frac{1}{D} \sum_{d=1}^D \left(\hat{\theta}^{**}_{b,d,h}-{\theta}^{**}_{b,d,h}\right)^2, \ b = 1,\ldots, B.
\end{equation}
The value $\hat{h}$ may be selected among the set $\mathcal{H}$ as the value minimizing the finite approximation of (\ref{eqn:risk_bootstrap}):
\begin{equation}
\label{eqn:hmin_db}
 \hat{h} = \underset{h\in \mathcal{H}}{\argmin} \ \frac{1}{B}\sum_{b=1}^B \left[\hat{V}^{**}_{b, h} - \hat{V}^*_h \right]^2.
\end{equation}
The resulting bootstrap variance estimator is therefore $\hat{V}^*_{\hat{h}}$, that is, one of the $m$ first-level estimators computed initially. The question then arises as to how to choose the grid $\mathcal{H}=\{ h_1, h_2, \ldots h_m\}$, or rather the grid of constants $\mathcal{C}=\{ C_1, C_2, \ldots, C_m\}$, where $C_i = h_i\cdot n^{1/5}$, $i=1,\ldots, m$. A sensible approach when faced with a particular sample is to initialize the grid and to adjust it over several executions of the double bootstrap until a global minimum point of $\widehat{\mathrm{MSE}}\left(\hat{V}^*_h\right) = B^{-1}\sum_{b=1}^B \left[\hat{V}^{**}_{b, h} - \hat{V}^*_h \right]^2$ is found.

The computational cost associated with the double bootstrap simulation method is evidently high, with a number of operations of order $mBD$. To reduce the number of operations, we suggest generating a single vector of $N^*$ i.i.d.~observations $(\varepsilon_1^*, \varepsilon_2^*, \ldots, \varepsilon_{N^*}^*)'$,  where $\varepsilon^*_i \sim K$ and $N^*$ is the size of the unsmoothed pseudo-population $U^*$, which can then be used repeatedly to create each of the $m$ smoothed pseudo-populations $U^*_h$, $h \in \mathcal{H}$ at a given bootstrap iteration. Similarly, for a given iteration, the random selection of the indices in $U^*$ according to a sampling design $p(\cdot)$ can be performed only once, implying that the same subset of $U^*$ is used to form the $m$ different smoothed bootstrap samples $S^*_h$, $h \in \mathcal{H}$. Note that these two schemes may be further applied to the second-level of bootstrap simulation as well.

{In the case of stratified SRSWOR, where the smoothed UEQPS PPB algorithm is to be applied within each stratum $l$, $l=1, \ldots, L$, independently to yield overall $\hat{\theta}^*_h$ and ${\theta}^*_h$ bootstrap distributions as described in Section \ref{section:smoothedbootstrapsurvey}, we suggest performing selection of the optimal constant $C$ using an overall grid of constants $\mathcal{C} =\{C_1, C_2, \ldots, C_m\}$ (as opposed to stratum-specific grids) with stratum-specific bandwidths in the form $h_{l, i} = C_i n_l^{-1/5}$, $i=1,\ldots, m$. This choice preserves the appropriate order for $h_l$ in each stratum while keeping the already high computational burden of the bootstrap bandwidth selection to a minimum. }

\subsection{Plug-in Method}
\label{subsection:plug-in}
We now proceed with describing a computationally simpler bandwidth selection method {for the case of SRSWOR}, which follows very closely that of \cite{silverman1986density} in the context of density estimation. As the title suggests, it consists of plugging sample estimates of unknown quantities into the expression of the optimal bandwidth for quantile variance estimation. 

In the i.i.d.~setting, \cite{hall1989smoothing} showed that, under certain smoothness and boundedness conditions on $f_0$ and for a kernel function $k$, the asymptotic mean squared error of the smoothed bootstrap variance estimator of the $p$-th level sample quantile with respect to the asymptotic quantile variance is minimized by a bandwidth equal to $h_{\text{opt}}^k(\xi_p) = C_{\text{opt}}^k(\xi_p)n^{-1/5}$, where 
\begin{equation}
\label{eqn:copt_iid_boot}
C_{\text{opt}}^k(\xi_p) = \kappa_1^{-2/5}{\kappa_2}^{1/5} \left[ f_0(\xi_p)\right]^{1/5} \left[f_0''(\xi_p) - f_0'(\xi_p)^2f_0(\xi_p)^{-1} \right]^{-2/5},
\end{equation}
 and $\kappa_1= \int t^2k(t)dt$ and $\kappa_2= \int k^2(t)dt$.

As shown by \cite{chatterjee2011asymptotic}, the asymptotic quantile variance for the SRSWOR design under a superpopulation model is equal to that of the sample quantile computed on a sample of $n$ i.i.d. observations apart from the multiplicative factor $(1-f)$, the so-called finite population correction factor. This suggests a \textit{plug-in} approach for estimating the optimal bandwidth in the smoothed pseudo-population method under the SRSWOR design, using the optimal constant in (\ref{eqn:copt_iid_boot}) as the target constant.

To do so, we must first specify an explicit form for $f_0$ in (\ref{eqn:copt_iid_boot}) for there to be an expression to estimate from the survey data. Here we exclude a nonparametric approach, which would involve plugging in kernel density estimators of $f_0$, $f_0'$ and $f_0''$ evaluated at the estimator of the quantile and thus estimating three additional bandwidths. Following the approach of \cite{silverman1986density} for the i.i.d.~case, we assume that the units in $U$ are drawn from the superpopulation $N(\mu, \sigma^2)$ by posing $f_0(x)=\sigma^{-1}\phi((x-\mu)/\sigma)$, where $\phi$ is the standard normal probability density function. It is then straightforward to show that the optimal bandwidth to estimate the variance of the $p$-th level sample quantile for normally distributed data is given by
\begin{align}
\label{hopt_norm}
& h_{\text{opt,norm}}^k\left(z\right) \nonumber \\ &= \kappa_2^{-2/5}{\kappa_1}^{1/5}  \left[\frac{1}{\sigma}\phi\left(z\right)\right]^{1/5}\left[\frac{1}{\sigma}\phi''\left(z\right) - \left(\frac{1}{\sigma}\phi'\left(z\right)\right)^2\left(\frac{1}{\sigma}\phi\left(z\right)\right)^{-1} \right]^{-2/5}n^{-1/5},
\end{align}
where $z = (\tilde{\xi}_p - \mu)/\sigma$. A plug-in estimate $\hat{h}_{\text{plug-in}}^k$ of $h_{\text{opt,norm}}^k$ is obtained by substituting the unknown quantity $z$ with $\hat{z} = (\hat{\xi}_p - \hat{\mu})/\hat{\sigma}$ in (\ref{hopt_norm}), where $\hat{\mu} = \bar{y} = n^ {-1}\sum_{i \in S} y_i $ and $\hat{\sigma}^2 = s^2=  (n-1)^{-1} \sum_{i \in S} (y_i - \bar{y})^2$.

\section{Simulation Studies}
\label{sec5}
Simulation studies were conducted to assess the performance of smoothed pseudo-population bootstrap algorithm with regard to {mean squared error} estimation and confidence intervals in relation to its unsmoothed counterpart. We applied our methods to the estimation of two finite population quantiles, namely the median and the third quartile. {Three high entropy survey designs were considered, that is, SRSWOR, Poisson sampling and randomized systematic PPS sampling}. A Gaussian kernel $k$ was used throughout the simulations. All reported results are based on 2,000 simulation replicates.
\subsection{Generation of the Finite Populations}
\label{section:generation}
This study is based on the following scenarios for the population size, made of the crossing of two sample sizes and two sampling fractions: $U_1: \ n = 100, \ f= 7\%$; $U_2 : \ n = 100, \ f = 30\%$; $U_3 : \ n = 500, \ f = 7 \%$; $ U_4 : \ n = 500, \ f = 30 \%$. In each case, $U_i$ is a finite population of size $N=\floor{n/f}$.
Recall that in the case of Poisson sampling, the notation $n$ refers to the expected sample size, $\mathbb{E}_p[n_S]$. The different population sizes were considered in combination with several distributions $F_0$ to generate the finite populations. For each distribution scenario, we generated a single population $\tilde{U}$ by drawing $\tilde{N} = 7,142$ i.i.d.~observations from the given distribution, with $\tilde{N}$ being large enough to extract subsets of different sizes depending on the sample size and the sampling fraction under investigation. We then selected the first $N$ indices of $\tilde{U}$ for each population scenario. 

{We considered two scenarios for the underlying distribution common to all sampling designs: $F_0^{\mathrm{sym}}$ and $F_0^{\mathrm{asym}}$. 
The first population $F_0^{\mathrm{sym}}$ was generated according to the model
\begin{align}
\label{eqn:model_sym}
Y_i = \gamma X_i + \sigma \varepsilon_i, \quad \varepsilon_i \stackrel{\text{i.i.d.}}{\sim} \mathcal{N}(0,1), \quad X_i \stackrel{\text{i.i.d.}}{\sim} \chi^2_{\nu}, \quad i = 1,\ldots,\tilde{N},
\end{align}
where $\gamma = 0.6$, $\nu = 100$ and $\sigma = 12$. With this choice of parameters, the empirical correlation coefficient between the observations of the response variable $\bm{y} = (y_1, y_2, \ldots, y_{\tilde{N}})'$ and the regressor $\bm{x} = (x_1, x_2, \ldots, x_{\tilde{N}})'$ was $0.5795$, where $\tilde{N} = 7,142$ observations. For the proportional-to-size sampling designs, the first-order inclusion probabilities were set equal to $\pi_i = n x_i/(\sum_{j=1}^N x_j), \ i =1,\ldots, N$. Since $x_i > 0 \ \forall i$, it follows that the inclusion probabilities are guaranteed to be strictly positive. While the distribution of $Y_i$ is not exactly symmetric, with such a high number of degrees of freedom for $X_i$, the asymmetry is very small which is why we decided to refer to this population as $F_0^{\mathrm{sym}}$.}

{
A second population $F_0^{\mathrm{asym}}$ was generated according to the model
\begin{align}
\label{eqn:model_asym}
Y_i = X_i^{\beta}\cdot \varepsilon_i, \quad \log \varepsilon_i \stackrel{\text{i.i.d.}}{\sim} \mathcal{N}(0,\sigma^2_{\varepsilon}), \quad \log X_i \stackrel{\text{i.i.d.}}{\sim} \mathcal{N}(\mu_X,\sigma^2_X), \quad i = 1,\ldots,\tilde{N},
\end{align}
where $\beta = 0.5$, $\mu_X = 3$, $\sigma^2_X = 1$, and $\sigma^2_{\varepsilon} = 1.139$. 
For $i = 1, \ldots, \tilde{N}$, the model can be rewritten} {as $Y_i \sim \mathrm{Lognormal}(\beta \mu_X, \beta^2\sigma_X^2 + \sigma^2_{\varepsilon})$. This choice of parameters led to an empirical correlation coefficient between the outcome variable and the regressor of $0.3938$ in the largest population of size $\tilde{N} = 7,142$ observations. Again, for Poisson sampling and randomized systematic PPS designs, the first-order inclusion probabilities were set equal to $\pi_i = n x_i/(\sum_{j=1}^N x_j), \ i =1,\ldots, N$. For this skewed population, several observations did not meet the condition that $\pi_i$ must be smaller than 1, so the correction described in Section 3 was applied to the first-order inclusion probabilities in each of the four populations $U_k$. Also, note that the scenario $n=500$, $f=7$\% was excluded for randomized systematic PPS design due to the prohibitively high computation time associated with drawing a single sample. }

{A detailed discussion of results under the SRSWOR design can be found in Subsection \ref{subsection:resultssrswor}, while results under Poisson sampling and randomized systematic PPS sampling are presented in Sections \ref{appendix_poisson} and \ref{appendix:syspps} of the Appendix, respectively. Finally, Section \ref{appendix:srswor} of the Appendix displays previous simulation results under SRSWOR where two additional scenarios for the underlying distribution $F_0$ are considered: $\mathcal{N}(0,1)$ and $\mathrm{Lognormal}(0,1)$.}
\subsection{Parameters of the Bandwidth Selection Methods}
Let $\hat{V}_{\hat{h}}$ denote the bootstrap {mean squared error} estimator of $\hat{\xi}_p$. We considered three values for $\hat{h}$ (the third is only used with SRSWOR) : 
\begin{flushleft}
UNSMTHD : $\hat{h}=0$ (no smoothing);\\
BOOT : $\hat{h}$ corresponds to the value selected by double bootstrap given in (\ref{eqn:hmin_db}) on a fixed grid $\mathcal{H}(n)$ which depends on the scenario; \\
PLUG-IN : $\hat{h}$ is the plug-in estimator of the optimal bandwidth given in (\ref{hopt_norm}) under the assumption that the data are normally distributed and {only used with SRSWOR}.
\end{flushleft}

For the BOOT selection method, we considered bandwidths of the form $h=Cn^{-1/5}$, $C>0$, {which is the optimal order for quantile mean squared error estimation in the i.i.d setting}. The grid of bandwidths $\mathcal{H}(n)$ varied depending on the underlying distribution $F_0$ in conjunction with the quantile under consideration. In each case, they included 50 equidistant points and were constructed so as to cover the optimal multiplicative constant $C_{\text{opt}}^k(\xi_p)$ given in (\ref{eqn:copt_iid_boot}). With the choice of the standard Gaussian kernel $\phi$, this constant is equal to 
\begin{equation}
\label{eqn:copt}
C_{\text{opt}}^{\phi}(\xi_p) = \left[2\sqrt{\pi}\right]^{-1/5} \left[ f_0(\xi_p)\right]^{1/5} \left[f_0''(\xi_p) - f_0'(\xi_p)^2f_0(\xi_p)^{-1} \right]^{-2/5}.
\end{equation}
We distinguish this constant from $C_{\text{norm}}^{\phi}(\xi_p)$, which denotes the constant multiplying $n^{-1/5}$ in (\ref{hopt_norm}) that is estimated in the PLUG-IN method. With a standard Gaussian kernel, it is given by
\begin{equation}
\label{eqn:coptnorm}
C_{\text{norm}}^{\phi}(\xi_p) = \left[\frac{1}{\sigma}\phi\left(z\right)\right]^{1/5}\left[2\sqrt{\pi}\right]^{-1/5}\left[\frac{1}{\sigma}\phi''\left(z\right) - \left(\frac{1}{\sigma}\phi'\left(z\right)\right)^2\left(\frac{1}{\sigma}\phi\left(z\right)\right)^{-1} \right]^{-2/5},
\end{equation}
where $z = (\xi_p-\mu)/\sigma$. Naturally, if the underlying distribution is normal, (\ref{eqn:copt}) and (\ref{eqn:coptnorm}) coincide. We define the quantity (\ref{eqn:coptnorm}) to assess how much the target bandwidth in the PLUG-IN method differs from (\ref{eqn:copt}) in instances where the underlying distribution is not normal.
{See Table \ref{tab:grille_eassr} for the optimal constants along with the constant estimated by the plug-in method for the different superpopulations. Under $F_0^{\mathrm{sym}}$, the normal approximation of the $\chi^2$ distribution was used to calculate the optimal constants in Table \ref{tab:grille_eassr}, in which case the observations in the population follow the model given by (\ref{eqn:model_sym}). Given that $X_i$ is distributed according to $\chi^2_{100}$ independently of the Gaussian noise $\varepsilon_i$, it follows that $Y_i$ is approximately normally distributed with mean $\nu\gamma$ and variance $2\nu\gamma^2 + \sigma^2 $ for $i=1, \ldots, \tilde{N}$. Therefore, the optimal constant in (\ref{eqn:copt}) was evaluated using the normal density function and the numeric values $\gamma = 0.6$, $\nu = 100$ and $\sigma = 12$ mentioned above.}
\begin{table}[H]
\centering
\caption{Optimal constants used to construct the bandwidth grids $\mathcal{H}(n)$}
\begin{threeparttable}
\begin{tabular}{l c c }
\hline
Scenario  & $C_{\text{opt}}^{\phi}(\xi_p)$ & $C_{\text{norm}}^{\phi}(\xi_p)$ \\
\hline
${\xi}_{0.50}$, \ $F_0^{\mathrm{sym}}$ & 13.71 & 13.71 
\\
${\xi}_{0.75}$, \ $F_0^{\mathrm{sym}}$ & 14.35 & 14.35 
\\
${\xi}_{0.50}$, \ $F_0^{\mathrm{asym}}$ & 9.83 & 10.86 \\
${\xi}_{0.75}$, \ $F_0^{\mathrm{asym}}$ & 10.20 & 10.78 \\
\hline
\end{tabular}
\end{threeparttable}
\label{tab:grille_eassr}
\end{table}

 {Figures \ref{fig:instability_norm} and \ref{fig:instability_lnorm} show the range considered for the grids $\mathcal{H}(n)$ for the different quantiles and superpopulations under SRSWOR. For each quantile and superpopulation scenario, the same grids were used for all survey designs, including Poisson and randomized systematic PPS sampling, after verifying by simulation that a minimum mean squared error was achieved empirically for $\hat{V}^*_h$ over the range of fixed values in $\mathcal{H}(n)$.}
 
The smoothed pseudo-population bootstrap methods were applied to each simulated sample $S$ using $B =$ 1,000 bootstrap samples. For the BOOT selection methods, the second-level bootstrap {mean squared error} estimators given by (\ref{eqn:doublevar}) were each based on $D=$ 50 second-level bootstrap samples, a number of replicates deemed sufficient to yield a precise double bootstrap variance estimator \citep{efron1993introduction}. Although our focus is on {mean squared error} estimation, asymptotic and basic bootstrap confidence intervals were also calculated as described in Subsection \ref{subsection:veic} for the three values $\hat{h}$ with a confidence level of 95\%.  

\subsection{Measures of Performance}
\label{subsection:performance}
For all bootstrap methods under consideration, the {mean squared error} estimators were compared to the true mean squared error (MSE) of the quantile estimator $\hat{\xi}_p$, where $p=0.50, 0.75$. {For each quantile, population, and survey design scenario, the true MSE was first approximated by selecting 3,000 samples using the survey design under consideration}, and using 
$\text{MSE} = (3000)^{-1}\sum_{s=1}^{3000} (\hat{\xi}_p - \xi_p )^2$. The bootstrap {mean squared error} estimators $\hat{V}_{\hat{h}}$ were then compared on the basis of $R = $ 2,000 different simulated samples with regard to (i) relative bias, using $\text{Bias\%} = 100\times\text{MSE}^{-1}(R^{-1}\sum_{r=1}^R \hat{V}_{r, \hat{h}} - \text{MSE})$ and (ii) relative instability, using $\text{RRMSE\%} = 100\times \text{MSE}^{-1}(R^{-1}\sum_{r=1}^R (\hat{V}_{r, \hat{h}} - \text{MSE})^2)^{1/2} $. The relative instability $\text{RRMSE\%}$ of a smoothed bootstrap {mean squared error estimator} $\hat{V}_h$ based on a fixed bandwidth $h$ was evaluated for the bandwidths included in each grid $\mathcal{H}(n)$. The performance of the bootstrap methods was also assessed through the lower, upper and two-tail coverage error rates of the 95\% asymptotic and basic bootstrap confidence intervals given by $\text{L\%}=100\times R^{-1}\sum_{r=1}^R I(\xi_p < \xi_{p,L})$, $\text{U\%}=100\times R^{-1}\sum_{r=1}^R I(\xi_p > \xi_{p,U})$ and $\text{L\%} + \text{U\%}$ respectively, where $\xi_{p,L}$ and $\xi_{p,U}$ denote the lower and upper bounds of a confidence interval.  With $R$ = 2,000 replicates, provided that the true one-tail error rate is 2.5\%, the lower and upper coverage error rates will lie between 1.8\% and 3.2\% 95 out of 100 times. Similarly, assuming that the true two-tail error rate is 5\%, the acceptance region corresponding to significance level 0.05 is given by [4.0; 6.0].

\subsection{Results under Simple Random Sampling Without Replacement}
\label{subsection:resultssrswor}
The bias and relative instability obtained over the 2,000 simulations are reported in Table \ref{tab:rrmse_eassr} for each of the bandwidth selection approaches. As well, RRMSE\% values for all three methods considered are compared to a relative instability curve obtained for a fixed bandwidth grid in Figure \ref{fig:instability_norm} for the $F_0^{\mathrm{sym}}$ superpopulation and in Figure \ref{fig:instability_lnorm} for the $F_0^{\mathrm{asym}}$ superpopulation. Results show that, compared with the UNSMTHD method, substantial reductions in RRMSE\% can be achieved from the smoothed pseudo-population bootstrap approach with a data-driven selection of the bandwidth.

When the underlying distribution is close to symmetrical and approximately normal, the PLUG-IN approach, in addition to being computationally inexpensive, displays small bias when $n$ is large and is the most stable for all sampling fractions and sample sizes, whether it be for estimating the variance of the median or that of the third quartile. {Figure \ref{fig:instability_norm} reveals some discrepancies between the theoretical optimal bandwidth and the true minimizer of instability, which tend to increase with a larger sampling fraction.} 

However, as expected, the performance of the PLUG-IN approach is generally poor when the underlying density function is wrongly assumed to be normal, as shown by the results for a skewed superpopulation (Table \ref{tab:rrmse_eassr}). It is especially the case for the median and when the sample size is small or the sampling fraction is large, in which cases the large bias of the variance estimator drives the instability up. {For the same superpopulation, the PLUG-IN method generally yields a gain in stability over the unsmoothed method in the case of the third quartile, with the exception of the $n=500, f=30$\% scenario.} 

At the cost of higher computational time, the nonparametric bandwidth selection approach generally succeeds where the previous method fails. 
As shown in Table \ref{tab:rrmse_eassr}, for the $F_0^{\mathrm{asym}}$ superpopulation, the BOOT approach achieves RRMSE\% values that are either approximately equivalent or lower than that of the PLUG-IN selection method while generally outperforming the standard unsmoothed method. This is corroborated by the instability curves of the {mean squared error} estimator of the median in Figure \ref{fig:instability_lnorm}, where it can be seen that when there is a higher potential for improvement in terms of stability, the RRMSE\% value of the BOOT method is closer to the true minimum compared with the PLUG-IN and UNSMTHD methods. {However, Figure \ref{fig:instability_lnorm} also shows that in the case of the third quartile under the $F_0^{\mathrm{asym}}$ superpopulation for the larger sampling fraction ($f=30$\%), two local minimums can be observed in the true RRMSE\% curve and the BOOT selection method appears to select values that strike a compromise between the two. In general, more difficulties are experienced in selecting the optimal bandwidth for larger sampling fractions and smaller population sizes, which could be explained by a greater departure from the i.i.d.~setting. That being said, Figure \ref{fig:instability_lnorm2} in Section \ref{appendix:srswor} of the Appendix shows that for the $\mathrm{Lognormal}(0,1)$ populations considered, smoothing always provides an advantage over not smoothing in the case of $\hat{V}_{\hat{h}}\left({\hat{\xi}_{0.75}}\right)$.}

When the underlying distribution is close to symmetric, although the PLUG-IN method is more efficient, the smoothed {mean squared error} estimator based on the BOOT selection approach always outperforms its unsmoothed counterpart. In general, as shown by the boxplots of Figures \ref{fig:instability_norm} and \ref{fig:instability_lnorm}, the distribution of the bandwidth selected by double bootstrap has a greater interquartile range than that of the plug-in bandwidth estimator. 
\begin{table}[h]
{
\caption{\small Bias and instability of mean squared error estimators under simple random sampling without replacement.}
\centering
\footnotesize
\setlength\tabcolsep{2pt}
\resizebox{\textwidth}{!}{%
\begin{tabular*}{\textwidth}{@{\extracolsep{\fill}}*{9}{c}}
  \toprule
  & \multicolumn{4}{c}{$F_0^{\mathrm{sym}}$} & \multicolumn{4}{c}{$F_0^{\mathrm{asym}}$}\\
  \cmidrule(l){2-5} \cmidrule(l){6-9}
   &\multicolumn{2}{c}{$\mathrm{Var}_p(\hat{\xi}_{0.50})$} & \multicolumn{2}{c}{$\mathrm{Var}_p(\hat{\xi}_{0.75})$} & \multicolumn{2}{c}{$\mathrm{Var}_p(\hat{\xi}_{0.50})$} & \multicolumn{2}{c}{$\mathrm{Var}_p(\hat{\xi}_{0.75})$}\\
    \cmidrule(l){2-3} \cmidrule(l){4-5} \cmidrule(l){6-7} \cmidrule(l){8-9}
Method& Bias\% & RRMSE\% & Bias\% & RRMSE\% & Bias\% & RRMSE\% & Bias\% & RRMSE\% \\ 
  \midrule
 \multicolumn{4}{l}{ (i) $n=100$, $f=7$\%, $N = 1~428 $}\\
 \addlinespace
UNSMTHD  & 15.0 & 56.1 & 10.6 & 51.1 & 10.9 & 55.2 & 13.9 & 73.4 \\ 
BOOT & 26.9 & 41.8 & 16.8 & 34.8 & 8.4 & 43.0 & -1.3 & 43.8 \\ 
PLUG-IN & 20.8 & 33.8 & 10.3 & 27.1 & 81.8 & 110.0 & -13.9 & 36.3 \\  
 \addlinespace
  \multicolumn{4}{l}{(ii) $n=100$, $f=30$\%, $N = 333$}\\
   \addlinespace
  UNSMTHD & 16.8 & 54.9 & 4.8 & 47.8 & 8.4 & 48.5 & -0.1 & 46.7 \\
  BOOT & 32.1 & 44.1 & -6.4 & 21.8 & 13.6 & 47.1 & -31.3 & 39.9 \\
  PLUG-IN & 27.3 & 37.3 & -9.0 & 20.3 & 97.1 & 113.3 & -39.0 & 44.0 \\  
   \addlinespace
 \multicolumn{4}{l}{(iii) $n=500$, $f=7$\%, $N = 7~142$}\\
  \addlinespace
 UNSMTHD & 2.9 & 32.5 & 10.8 & 39.4 & 2.8 & 33.8 & 10.2 & 43.0 \\ 
 BOOT & 4.2 & 16.8 & 16.1 & 25.9 & -5.3 & 16.7 & -1.7 & 17.4 \\ 
  PLUG-IN & 4.4 & 13.8 & 17.2 & 23.2 & 32.9 & 46.0 & -2.9 & 17.3 \\ 
   \addlinespace
 \multicolumn{4}{l}{(iv) $n=500$, $f=30$\%, $N = 1~666$}\\
  \addlinespace
  UNSMTHD & 2.9 & 31.9 & 4.5 & 34.7 & 8.5 & 35.8 & -7.2 & 34.0 \\  
  BOOT & -4.1 & 16.0 & 12.4 & 22.4 & -0.4 & 14.2 & -37.5 & 38.8 \\ 
  PLUG-IN & -0.5 & 11.6 & 14.1 & 20.1 & 38.6 & 43.8 & -38.5 & 39.9 \\ 
   \bottomrule
\end{tabular*}}
\label{tab:rrmse_eassr}
}
\end{table}

\begin{figure}
\centering
\small{Instability curves for $\hat{V}_{\hat{h}}\left(\hat{\xi}_{0.50}\right)$ with underlying distribution $F_0^{\mathrm{sym}}$}\\
\includegraphics[scale=0.5]{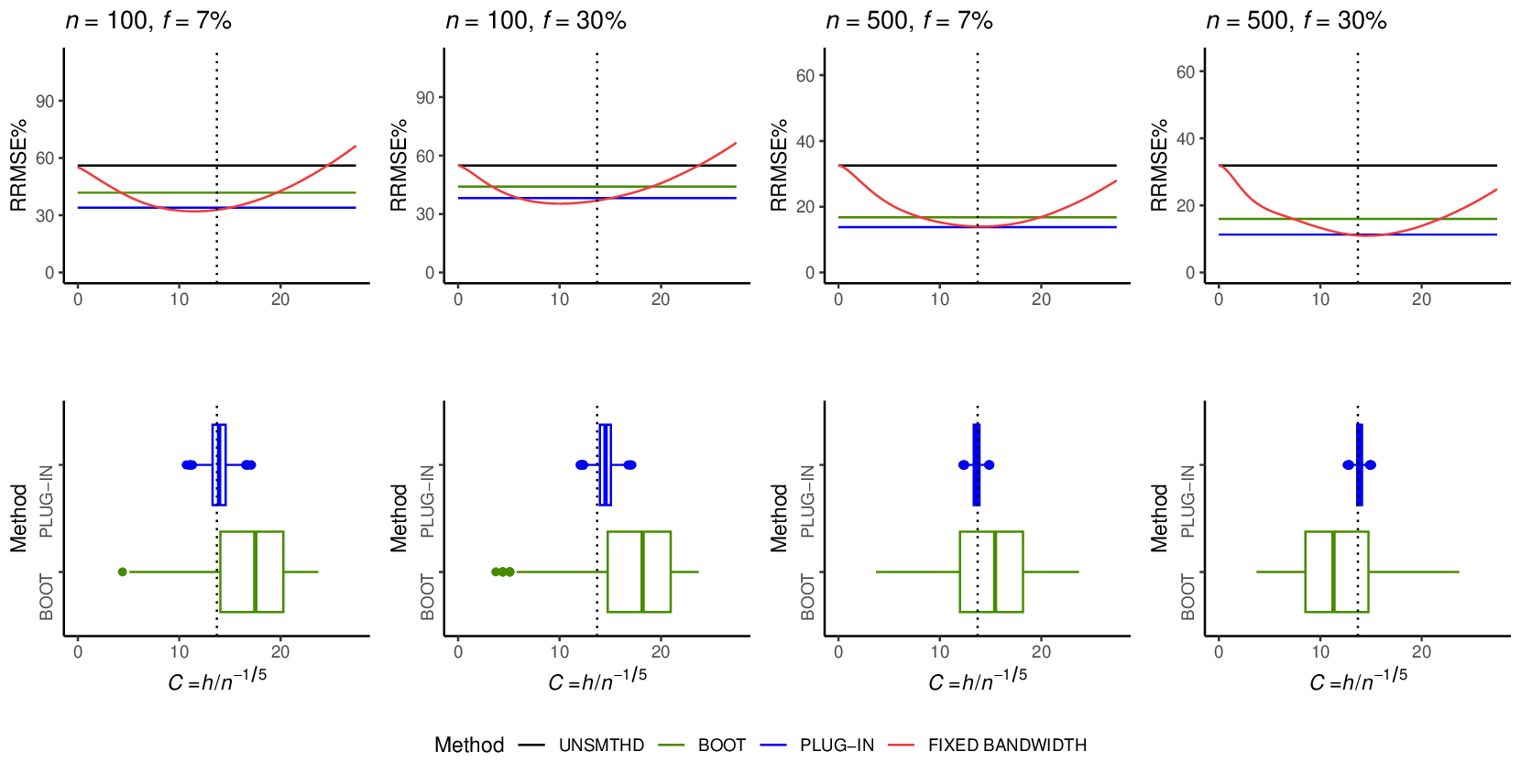}\\
\small{Instability curves for $\hat{V}_{\hat{h}}\left(\hat{\xi}_{0.75}\right)$ with underlying distribution $F_0^{\text sym}$}\\
\includegraphics[scale=0.5]{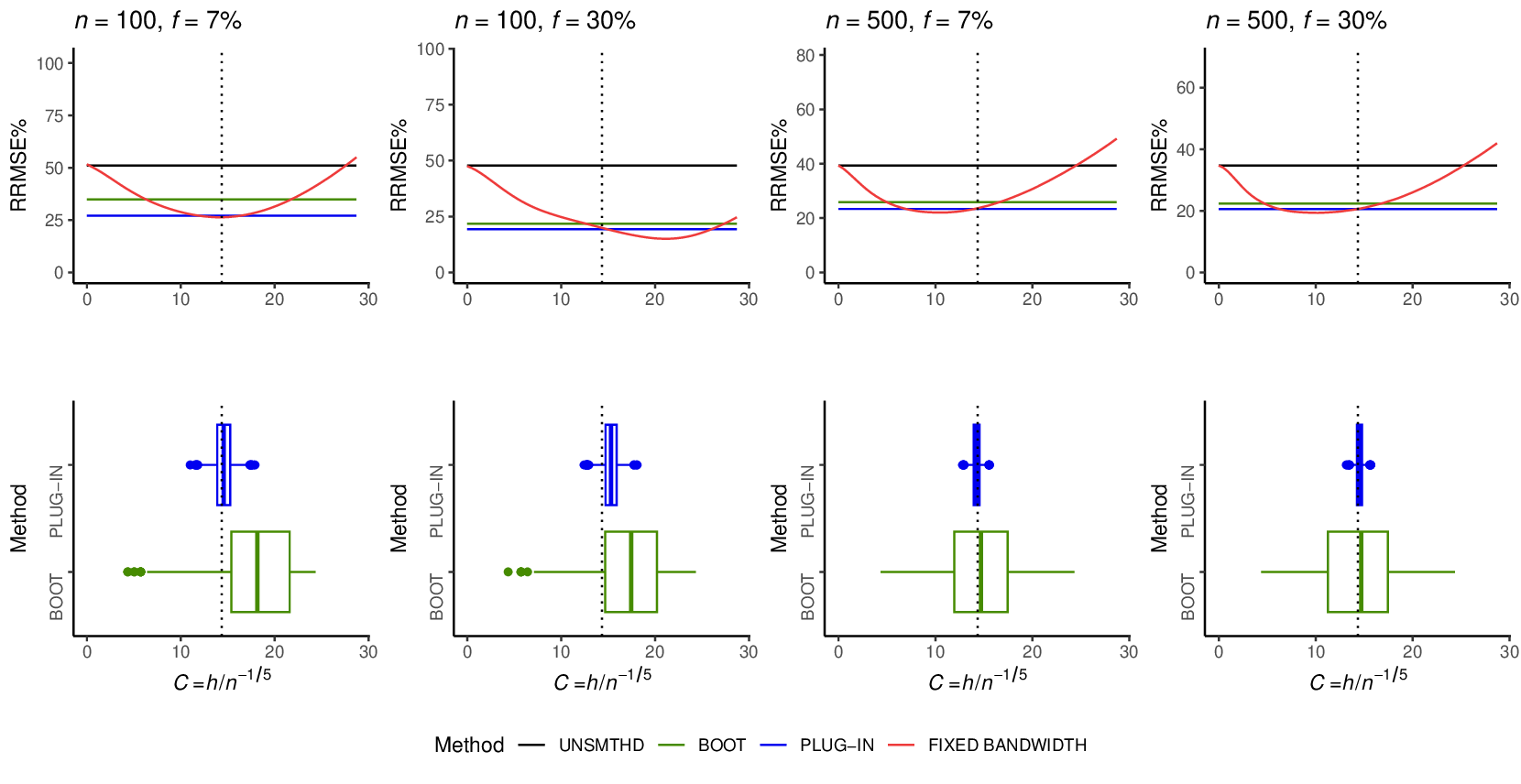}\\
\caption[Instability (RRMSE\%) curves of bootstrap mean squared error estimators for four approaches and boxplots of selected bandwidths $\hat{C} = \hat{h}/n^{-1/5}$ for the BOOT and PLUG-IN approaches across $R = $ 2,000 simulated SRSWOR samples from a single finite population with ($n$, $f$) = (100, 0.07), (100, 0.30), (500, 0.07), (500, 0.30) and underlying distribution $F_0^{\text sym}$]{ Instability (RRMSE\%) curves of bootstrap mean squared error estimators for four approaches and boxplots of selected bandwidths $\hat{C} = \hat{h}/n^{-1/5}$ for the BOOT and PLUG-IN approaches across $R = $ 2,000 simulated SRSWOR samples from a single finite population with ($n$, $f$) = (100, 0.07), (100, 0.30), (500, 0.07), (500, 0.30) and underlying distribution $F_0^{\text sym}$. The red curve in the plots corresponds to the RRMSE\% of $\hat{V}_h$, where $h = C \cdot n^{-1/5}$ and $C$ is a known (fixed) constant. The solid black line is the RRMSE\% of the unsmoothed (standard) bootstrap method ($\hat{h} = 0$). The solid blue and green lines pertain to the PLUG-IN and BOOT bandwidth selection methods, respectively. The vertical dotted line corresponds to the value of the optimal bandwidth constant given in (\ref{eqn:copt}) for the standard normal distribution. }
\label{fig:instability_norm}
\end{figure}

\begin{figure}
\centering
\small{Instability curves for $\hat{V}_{\hat{h}}\left(\hat{\xi}_{0.50}\right)$ with underlying distribution $F_0^{\mathrm{asym}}$}\\
\includegraphics[scale=0.5]{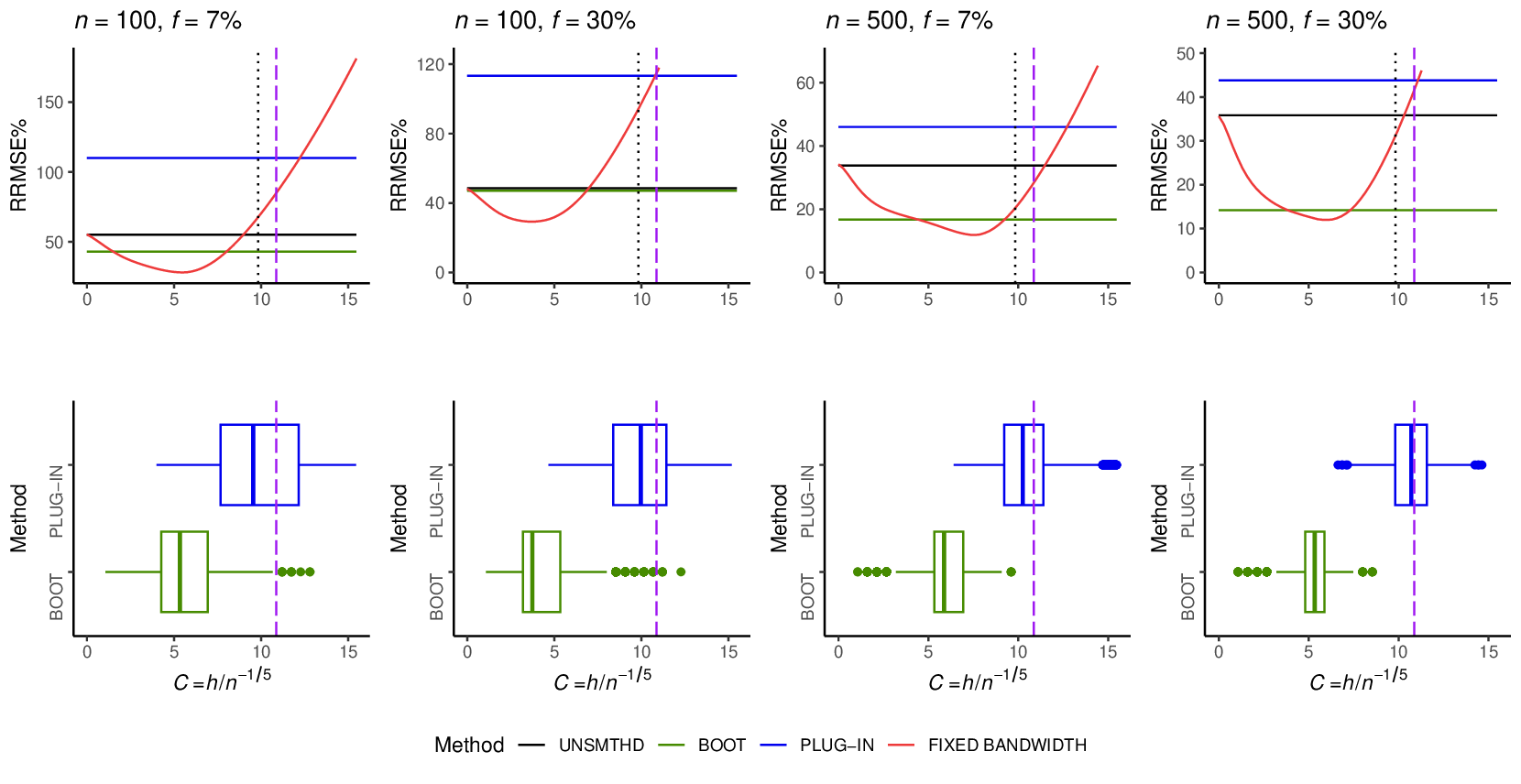}\\
\small{Instability curves for $\hat{V}_{\hat{h}}\left(\hat{\xi}_{0.75}\right)$ with underlying distribution $F_0^{\mathrm{asym}}$}\\
\includegraphics[scale=0.5]{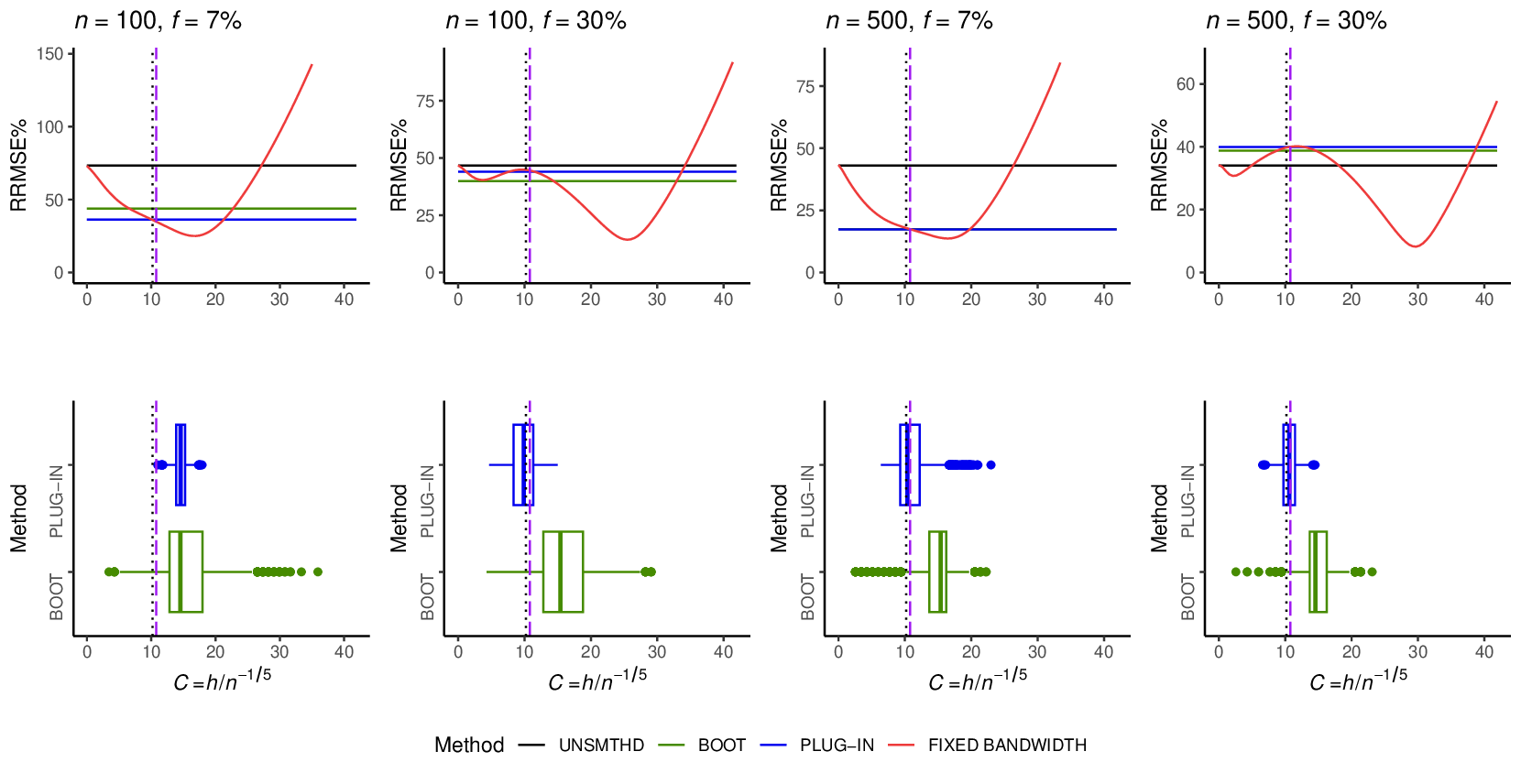}\\
\caption[Instability (RRMSE\%) curves of bootstrap mean squared error estimators for four approaches and boxplots of selected bandwidths $\hat{C} = \hat{h}/n^{-1/5}$ for the BOOT and PLUG-IN approaches across $R = $ 2,000 simulated SRSWOR samples from a single finite population with ($n$, $f$) = (100, 0.07), (100, 0.30), (500, 0.07), (500, 0.30) and underlying distribution $F_0^{\mathrm{asym}}$]{ Instability (RRMSE\%) curves of bootstrap mean squared error estimators for four approaches and boxplots of selected bandwidths $\hat{C} = \hat{h}/n^{-1/5}$ for the BOOT and PLUG-IN approaches across $R = $ 2,000 simulated SRSWOR samples from a single finite population with ($n$, $f$) = (100, 0.07), (100, 0.30), (500, 0.07), (500, 0.30) and underlying distribution $F_0^{\mathrm{asym}}$. The red curve in the plots corresponds to the RRMSE\% of $\hat{V}_h$, where $h = C \cdot n^{-1/5}$ and $C$ is a known (fixed) constant. The solid black line is the RRMSE\% of the unsmoothed (standard) bootstrap variance estimator ($\hat{h} = 0$). The solid blue and green lines pertain to the PLUG-IN and BOOT bandwidth selection methods, respectively. The vertical dotted black line corresponds to the value of the optimal bandwidth constant given in  (\ref{eqn:copt}) for the distribution $F_0^{\mathrm{asym}}$. The vertical dashed purple line is the optimal bandwidth constant for normally distributed observations given in (\ref{eqn:coptnorm}). {Note the near overlap in RRMSE\% for the PLUG-IN and BOOT methods in the scenarios $n=500$, $f=7\%$ and $n=500$, $f=30\%$ for $\hat{V}_{\hat{h}}\left(\hat{\xi}_{0.75} \right)$.}}
\label{fig:instability_lnorm}
\end{figure}

Table \ref{tab:asymptotic_srs} shows the one-tail and two-tail observed error rates of 95\% asymptotic bootstrap confidence intervals for nominal level. While it is difficult to draw any clear-cut conclusion, the unsmoothed method performs fairly well despite a lackluster performance for the lognormal superpopulation. One-tail errors are not well tracked in general. Overall, the two-tail coverage associated with smoothed methods tends to be higher than the nominal level of 95\%. This is consistent with the fact that the smoothed asymptotic bootstrap confidence intervals also tend to be longer than their unsmoothed counterpart (results not shown).

Coverage error rates of $95$\% level basic bootstrap confidence intervals are summarized in Table \ref{tab:basic_srs}. Here, the situation is somewhat reversed, in the sense that the unsmoothed method performs very poorly, with very high two-tail errors in almost all instances. On the other hand, smoothed basic confidence intervals are conservative at worst. Moreover, the observed two-tail error rate of BOOT confidence intervals falls more frequently within the acceptance region of $[4.0;6.0]$ compared with the other methods.
\begin{table}[h]
{
\caption{ Coverage error rates of $95$\% level asymptotic bootstrap confidence intervals under simple random sampling without replacement.}
\centering
\footnotesize
\setlength\tabcolsep{2pt}
\begin{threeparttable}
\resizebox{\textwidth}{!}{
\begin{tabular*}{\textwidth}{@{\extracolsep{\fill}}*{14}{c}}
\toprule
  & \multicolumn{6}{c}{$F_0^{\mathrm{sym}}$} & \multicolumn{6}{c}{$F_0^{\mathrm{asym}}$}\\
  \cmidrule(l){2-7} \cmidrule(l){8-13}
   &\multicolumn{3}{c}{${\xi}_{0.50}$} & \multicolumn{3}{c}{${\xi}_{0.75}$} & \multicolumn{3}{c}{${\xi}_{0.50}$} & \multicolumn{3}{c}{${\xi}_{0.75}$}\\
    \cmidrule(l){2-4} \cmidrule(l){5-7} \cmidrule(l){8-10} \cmidrule(l){11-13}
 & $\text{L\%}$ & $\text{U\%}$ & $\text{L+U\%}$ & $\text{L\%}$ & $\text{U\%}$ & $\text{L+U\%}$ & $\text{L\%}$ & $\text{U\%}$ & $\text{L+U\%}$ & $\text{L\%}$ & $\text{U\%}$ & $\text{L+U\%}$ \\ 
\midrule
 \multicolumn{13}{l}{ (i) $n=100$, $f=7$\%, $N = 1~428 $}\\
\addlinespace
UNSMTHD & 1.7 & 4.3 & 6.0 & 1.8 & 4.0 & 5.9 & 3.2 & 3.4 & 6.6 & 2.1 & 6.4 & 8.5 \\ 
BOOT & 0.9 & 1.8 & 2.6 & 0.9 & 3.6 & 4.5 & 2.1 & 2.6 & 4.7 & 2.6 & 3.0 & 5.7 \\
 PLUG-IN & 1.1 & 1.8 & 2.9 & 0.8 & 3.7 & 4.5 & 0.7 & 0.9 & 1.6 & 3.2 & 4.1 & 7.3 \\ 
 \addlinespace
   \multicolumn{13}{l}{(ii) $n=100$, $f=30$\%, $N = 333$}\\
   \addlinespace
    UNSMTHD & 1.7 & 2.9 & 4.5 & 2.1 & 5.1 & 7.3 & 2.7 & 3.5 & 6.2 & 1.4 & 9.1 & 10.5 \\ 
  BOOT & 0.4 & 2.9 & 3.4 & 0.9 & 6.6 & 7.5 & 1.1 & 3.5 & 4.7 & 1.2 & 14.8 & 16.1 \\ 
  PLUG-IN & 0.4 & 3.2 & 3.6 & 0.6 & 6.8 & 7.3 & 0.1 & 0.7 & 0.8 & 1.4 & 16.7 & 18.1 \\  
  \addlinespace
   \multicolumn{13}{l}{(iii) $n=500$, $f=7$\%, $N = 7~142$}\\
   \addlinespace
UNSMTHD & 3.5 & 2.9 & 6.5 & 1.8 & 2.5 & 4.3 & 2.8 & 4.2 & 7.0 & 2.4 & 3.5 & 5.8 \\
 BOOT & 2.2 & 3.1 & 5.4 & 0.8 & 2.7 & 3.5 & 2.7 & 3.7 & 6.4 & 2.8 & 3.3 & 6.0 \\
  PLUG-IN & 2.2 & 3.2 & 5.5 & 0.7 & 2.8 & 3.5 & 1.5 & 1.6 & 3.1 & 2.7 & 3.2 & 5.9 \\
  \addlinespace
\multicolumn{13}{l}{(iv) $n=500$, $f=30$\%, $N = 1~666$}\\
\addlinespace
    UNSMTHD & 1.5 & 4.2 & 5.7 & 1.8 & 4.6 & 6.5 & 3.1 & 2.9 & 6.0 & 2.6 & 7.3 & 10.0 \\ 
  BOOT & 1.2 & 5.6 & 6.9 & 1.0 & 5.0 & 5.9 & 3.8 & 2.9 & 6.8 & 4.3 & 8.7 & 13.1 \\ 
  PLUG-IN & 1.1 & 4.9 & 6.0 & 0.9 & 4.6 & 5.5 & 1.5 & 1.4 & 2.9 & 4.2 & 9.2 & 13.4 \\ 
   \bottomrule
\end{tabular*}}
\begin{tablenotes}
\footnotesize
    \item The 95\% confidence intervals for the two-tail 5.0\% and the one-tail 2.5\% nominal error rates are [4.0; 6.0]\% and [1.8; 3.2]\%, respectively.
  \end{tablenotes}
\end{threeparttable}
\label{tab:asymptotic_srs}
}
\end{table}

\begin{table}[h]
{
\caption{Coverage error rates of $95$\% level basic bootstrap confidence intervals under simple random sampling without replacement.}
\centering
\footnotesize
\begin{threeparttable}
\begin{tabular*}{\textwidth}{@{\extracolsep{\fill}}*{14}{c}}
\toprule
  & \multicolumn{6}{c}{$F_0^{\mathrm{sym}}$} & \multicolumn{6}{c}{$F_0^{\mathrm{asym}}$}\\
  \cline{2-7} \cline{8-13}
   &\multicolumn{3}{c}{${\xi}_{0.50}$} & \multicolumn{3}{c}{${\xi}_{0.75}$} & \multicolumn{3}{c}{${\xi}_{0.50}$} & \multicolumn{3}{c}{${\xi}_{0.75}$}\\
    \cline{2-4} \cline{5-7} \cline{8-10} \cline{11-13}
 & $\text{L\%}$ & $\text{U\%}$ & $\text{L+U\%}$ & $\text{L\%}$ & $\text{U\%}$ & $\text{L+U\%}$ & $\text{L\%}$ & $\text{U\%}$ & $\text{L+U\%}$ & $\text{L\%}$ & $\text{U\%}$ & $\text{L+U\%}$ \\ 
\midrule
 \multicolumn{13}{l}{ (i) $n=100$, $f=7$\%, $N = 1~428 $}\\
 \addlinespace
UNSMTHD & 3.6 & 10.2 & 13.8 & 4.0 & 7.3 & 11.2 & 4.8 & 8.9 & 13.7 & 5.2 & 13.6 & 18.9 \\
 BOOT & 1.1 & 2.2 & 3.3 & 1.1 & 3.2 & 4.4 & 1.7 & 3.9 & 5.5 & 2.2 & 3.9 & 6.1 \\  
  PLUG-IN & 1.2 & 2.2 & 3.5 & 1.7 & 3.9 & 5.5 & 0.5 & 1.1 & 1.7 & 2.5 & 5.5 & 8.0 \\ 
   \addlinespace
   \multicolumn{13}{l}{(ii) $n=100$, $f=30$\%, $N = 333$}\\
    \addlinespace
  UNSMTHD & 3.5 & 6.3 & 9.8 & 7.0 & 9.8 & 16.8 & 4.3 & 6.2 & 10.4 & 4.9 & 22.1 & 27.0 \\ 
  BOOT & 0.6 & 3.0 & 3.6 & 1.4 & 5.3 & 6.7 & 0.9 & 4.7 & 5.6 & 0.8 & 15.1 & 15.8 \\ 
  PLUG-IN & 0.4 & 3.4 & 3.9 & 1.2 & 5.8 & 7.0 & $<0.1$ & 0.9 & 0.9 & 0.9 & 18.8 & 19.8 \\  
   \addlinespace
   \multicolumn{13}{l}{(iii) $n=500$, $f=7$\%, $N = 7~142$}\\
    \addlinespace
 UNSMTHD & 5.5 & 4.5 & 10.1 & 4.1 & 3.7 & 7.8 & 3.8 & 7.8 & 11.7 & 4.0 & 5.4 & 9.4 \\ 
  BOOT & 2.5 & 3.2 & 5.7 & 1.5 & 2.5 & 4.0 & 2.1 & 4.5 & 6.6 & 2.5 & 3.5 & 6.0 \\ 
 PLUG-IN & 2.5 & 3.3 & 5.8 & 0.9 & 2.8 & 3.6 & 1.5 & 1.7 & 3.1 & 2.4 & 3.8 & 6.1 \\ 
  \addlinespace
\multicolumn{13}{l}{(iv) $n=500$, $f=30$\%, $N = 1~666$}\\
 \addlinespace
 UNSMTHD & 2.1 & 7.4 & 9.4 & 2.9 & 6.7 & 9.6 & 3.8 & 4.7 & 8.5 & 6.6 & 14.0 & 20.5 \\  
  BOOT  & 1.4 & 5.6 & 7.0 & 1.2 & 4.4 & 5.7 & 3.3 & 3.5 & 6.8 & 3.7 & 9.0 & 12.8 \\ 
 PLUG-IN & 1.1 & 4.9 & 6.0 & 1.1 & 4.0 & 5.1 & 1.4 & 1.5 & 2.9 & 3.4 & 10.5 & 13.9 \\ 
   \hline
\end{tabular*}
\begin{tablenotes}
\footnotesize
    \item The 95\% confidence intervals for the two-tail 5.0\% and the one-tail 2.5\% nominal error rates are [4.0; 6.0]\% and [1.8; 3.2]\%, respectively.
  \end{tablenotes}
\end{threeparttable}
\label{tab:basic_srs}
}
\end{table}

Relative instability of the mean squared error estimators and empirical coverage error rates under Poisson sampling are presented in Tables \ref{tab:rrmse_poisson}, \ref{tab:asymptotic_poisson}, and \ref{tab:basic_poisson} in Section \ref{appendix_poisson} of the Appendix. Under this sampling scheme, results suggest higher efficiency for the smoothed bootstrap variance estimator with the BOOT selection method as compared to no smoothing, regardless of the underlying data distribution. Again, smoothed asymptotic confidence intervals tend to be too conservative. However, while unsmoothed basic bootstrap confidence intervals are associated with very large one-tail and two-tail coverage errors, the smoothed basic intervals generally show good performance. {Similar conclusions can be drawn for the randomized systematic PPS design, for which instability and empirical coverage results are displayed in Tables \ref{tab:rrmse_syspps},\ref{tab:asymptotic_syspps}, and \ref{tab:basic_syspps} in Section \ref{appendix:syspps} of the Appendix.

To conclude this section, it is worth emphasizing that in our simulation setup, we generate a single finite population, such that results only account for variability due to the design. The true MSE of $\hat{\xi}_p$ we approximated in Subsection \ref{subsection:performance} corresponds, in fact, to a design MSE, that is to say, $\mathbb{E}_p[(\hat{\xi}_p - \xi_p)^2]$. However, we could consider an additional source of variation due to the model, in conformity with the superpopulation framework of \cite{isaki1982survey}. This would entail sampling a new finite population for each simulation replicate, yielding the target MSE $\mathbb{E}_0\mathbb{E}_p[(\hat{\xi}_p - \xi_p)^2]$, where $\mathbb{E}_0$ denotes the expectation with respect to the superpopulation model $F_0$. By allowing the finite population to vary randomly, we would be able to confirm the optimality of the values $C_{\text{opt}}^{\phi}(\xi_p)$ determined theoretically in (\ref{eqn:copt})  \citep{mcnealis2019estimateur}. Numerical discrepancies between the theoretical values and the observed minimums in Figures \ref{fig:instability_norm} and \ref{fig:instability_lnorm} are contingent upon the single finite population we generated.

\section{Concluding Remarks}
\label{sec6}
Whether it be for variance, coefficient of variation or confidence interval estimation, statistical agencies routinely rely upon bootstrap resampling methods, especially when faced with nonlinear functions of means such as quantiles.  In that regard, pseudo-population bootstrap methods are particularly attractive for they inherit many of the features of the sampling designs that they emulate. Nevertheless, deficiencies remain when the functional of interest depends on local properties of the underlying distribution, hence the idea of smoothing the bootstrap. {In this work, we demonstrated the applicability of the smoothed bootstrap within the context of survey sampling and described a smoothed pseudo-population bootstrap algorithm for unequal single-stage probability designs.} Given that the proposed bootstrap estimators are indexed by a smoothing parameter, we proposed a distribution-free approach to select the bandwidth, a double pseudo-population bootstrap procedure, which entails minimizing a bootstrap estimate of a risk function. We additionally described a less computationally intensive plug-in approach for bandwidth selection, which makes strong assumptions about the underlying data distribution. In light of the work of \cite{hall1989smoothing} in the i.i.d.~context, it is not surprising that our approach does well for the {mean squared error} of sample quantiles as compared with the unsmoothed method. In instances where smoothing is theoretically advantageous, it leads to significant improvement in a mean squared error sense.

As with all proposals, ours also comes with certain noteworthy limitations. {First, smoothing primarily improves stability rather than bias and its benefits should be weighed against any increase in bias. Smoothing does not always reduce relative bias and coverage improvements can be modest.} The plug-in method for bandwidth selection is restricted to SRSWOR and makes assumptions about the underlying form of the distribution, which, as one could argue, goes against the spirit of the nonparametric bootstrap for functional estimation. On the other hand, while the double bootstrap selection method can be used to estimate the {mean squared error} of a sample quantile under a variety of high entropy sampling schemes, the higher flexibility associated with this bandwidth selection approach comes at a higher computational cost. It also requires careful selection of a grid of candidate bandwidths. {Even though confidence interval coverage was not the primary focus of this paper, it is worth mentioning that smoothing may not improve coverage properties of asymptotic intervals, regardless of the bandwidth selection method being used.} 
{It must be noted that the smoothing parameter was selected to optimize the estimation of the mean squared error, not the coverage of the confidence intervals. Empirical coverage rates could be improved if the risk function to minimize in the bandwidth selection procedure was some function of the coverage error of the smoothed confidence interval, instead of the mean squared error of the smoothed variance estimator.} Preliminary work using the check loss function for asymmetric measurement of coverage error \citep{calonico2022coverage} has shown promising results, although other loss functions would be worth investigating {\citep{mcnealis2019estimateur}}. Further research on this matter is required. 

{As mentioned previously, the extension of the smoothed pseudo-population bootstrap approach to a stratified SRSWOR design would be straightforward in that it would entail carrying out the algorithm independently within strata with possibly stratum-specific bandwidths.} Note that several generalizations of pseudo-population procedures also exist for multistage designs; see, for instance, \cite{sitter1992comparing} and \cite{chauvet2007bootstrap} for two-stage cluster sampling and \cite{saigo2010comparing} for stratified three-stage sampling. In the case of multistage designs where SRSWOR is used a every stage and sampling fractions are high, a Bernoulli-type bootstrap was proposed by \cite{funaoka2006bernoulli} which can accommodate any number of stages. Future work could involve empirical evaluation of smoothing in multistage bootstrap methods. 

{While design weights were used in the quantile estimation procedure, these could be replaced with calibrated weights in the smoothed pseudo-population bootstrap procedure, enabling adjustment for auxiliary information and improving the precision of point estimators. For linear parameters, such as} {means and population totals, GREG weights can improve efficiency by adjusting sampling weights so that weighted estimates of auxiliary variables match known population totals \citep{deville1992calibration, sardnal1992model}. Calibration weights in the like of GREG weights have been generalized to other statistics than means and population totals. For instance, several quantile estimators were also developed in the calibration framework, such as the estimators based on quantile calibration constraints proposed by \cite{harms2006calibration}, or the more recently proposed estimators based on joint calibration of population totals and quantiles \citep{berkesewicz2023note}. Future work could entail assessing the improvement in bootstrap variance estimation when combining calibration and smoothing when estimating finite population quantiles. Note that this would entail adjusting the calibration constraints to each smoothed pseudo-population generated during bootstrap sampling since the pseudo-population of the auxiliary variable would also change for each bootstrap iteration.}

{Lastly, another possible avenue for future research could be to extend the smoothed pseudo-population bootstrap method to accommodate missing data, particularly under negligible sampling fractions. \cite{chen2019pseudo} developed pseudo-population bootstrap methods tailored for imputed survey data and derived variance estimators with respect to the nonresponse model or the imputation model inferential approaches. Smoothing the pseudo-population values of the study variable in the bootstrap methods of \cite{chen2019pseudo} could potentially improve variance estimators for quantiles when dealing with item nonresponse.}

\section*{Acknowledgements}

The authors appreciate the helpful comments of the two anonymous referees and the associate editor. This research was enabled in part by support provided by Calcul Québec (\url{https://www.calculquebec.ca/}) and the Digital Research Alliance of Canada (\url{https://alliancecan.ca/en}).

\section*{Disclosure Statement}

Vanessa McNealis is supported by a postdoctoral fellowship from Natural Sciences and Engineering Research Council of Canada (NSERC). Christian Léger acknowledges support from an NSERC Discovery Grant [RGPIN-2016-05686]. The authors have no conflicts of interest to disclose.

\section*{Data Availability}

The simulation code and synthetic datasets are available from the corresponding
author on reasonable request. Sample code can be found on github:\\ \url{https://github.com/vanessamcnealis/Smoothed-Pseudo-Population-Bootstrap}.


\newpage
\bibliographystyle{apacite} 
\bibliography{Sample}


\newpage
\appendix
\setcounter{table}{0}
\setcounter{figure}{0}
\setcounter{equation}{0}
\setcounter{algorithm}{0}
\renewcommand{\thefigure}{S\arabic{figure}}
\renewcommand{\thetable}{S\arabic{table}}
\renewcommand{\theequation}{S\arabic{equation}}
\renewcommand{\thealgorithm}{S\arabic{algorithm}}
\section{{Results under Poisson Sampling}}

\label{appendix_poisson}
In this subsection, the performance of the UNSMTHD and BOOT methods is assessed under Poisson sampling. Table \ref{tab:rrmse_poisson} summarizes the bias and the instability measures of the variance estimators for the various scenarios enumerated in Subsection \ref{section:generation}. As was the case for the $F_0^{\mathrm{sym}}$ superpopulation under SRSWOR, the smoothed bootstrap {mean squared error} estimator shows significant improvement in terms of stability over the standard pseudo-population bootstrap approach without any exception. In constrast with the case of SRSWOR, the bias of the {mean squared error} estimator does not vanish quickly as the expected sample size increases when $\hat{h}=0$ (UNSMTHD method).

Performance measures for the asymptotic and basic bootstrap confidence intervals are reported in Tables \ref{tab:asymptotic_poisson} and \ref{tab:basic_poisson}, respectively. Since asymptotic smoothed bootstrap confidence intervals generally tend to be too wide and thus too conservative, they bring little to no improvement over the asymptotic unsmoothed bootstrap confidence intervals, whose two-tail error is not significantly different than the nominal level in most cases. Again, conclusions are very different in the case of basic confidence intervals, for which no smoothing can lead to severe undercoverage. In this instance, the greater length of the smoothed basic bootstrap confidence intervals comes at an advantage, especially if we set the cost of undercoverage to be higher than that of overcoverage. The overall poor performance regarding one-tail errors that was observed for SRSWOR can also be noted here.

\begin{table}[H]
{
\caption{\small Bias and instability of mean squared error estimators under Poisson sampling.}
\label{tab:rrmse_poisson}
\centering
\footnotesize
\setlength\tabcolsep{2pt}
\resizebox{\textwidth}{!}{%
\begin{tabular*}{\textwidth}{@{\extracolsep{\fill}}*{9}{c}}
  \toprule
  & \multicolumn{4}{c}{$F_0^{\mathrm{sym}}$} & \multicolumn{4}{c}{$F_0^{\mathrm{asym}}$}\\
  \cmidrule(l){2-5} \cmidrule(l){6-9}
   &\multicolumn{2}{c}{$\mathrm{Var}_p(\hat{\xi}_{0.50})$} & \multicolumn{2}{c}{$\mathrm{Var}_p(\hat{\xi}_{0.75})$} & \multicolumn{2}{c}{$\mathrm{Var}_p(\hat{\xi}_{0.50})$} & \multicolumn{2}{c}{$\mathrm{Var}_p(\hat{\xi}_{0.75})$}\\
    \cmidrule(l){2-3} \cmidrule(l){4-5} \cmidrule(l){6-7} \cmidrule(l){8-9}
Method& Bias\% & RRMSE\% & Bias\% & RRMSE\% & Bias\% & RRMSE\% & Bias\% & RRMSE\% \\ 
  \midrule
 \multicolumn{4}{l}{ (i) $n=100$, $f=7$\%, $N = 1~428 $}\\
 \addlinespace
UNSMTHD & 14.7 & 58.2 & 21.7 & 63.7 & 20.1 & 83.0 & 18.8 & 114.4 \\ 
BOOT & 25.4 & 43.3 & 24.6 & 42.1 & 16.2 & 59.2 & -0.1 & 66.7 \\ 
 \addlinespace
  \multicolumn{4}{l}{(ii) $n=100$, $f=30$\%, $N = 333$}\\
   \addlinespace
  UNSMTHD & 19.6 & 59.9 & 13.2 & 55.4 & 19.2 & 87.8 & 0.1 & 71.0 \\ 
  BOOT & 33.0 & 46.8 & 1.4 & 23.9 & 21.1 & 72.2 & -20.9 & 47.0 \\ 
   \addlinespace
 \multicolumn{4}{l}{(iii) $n=500$, $f=7$\%, $N = 7~142$}\\
  \addlinespace
 UNSMTHD & 11.4 & 37.7 & 13.4 & 40.3 & 6.0 & 43.5 & 8.2 & 50.7 \\
 BOOT & 12.7 & 22.3 & 21.2 & 30.5 & -5.2 & 26.2 & -5.7 & 26.1 \\  
   \addlinespace
 \multicolumn{4}{l}{(iv) $n=500$, $f=30$\%, $N = 1~666$}\\
  \addlinespace
  UNSMTHD & 4.1 & 32.4 & 10.1 & 36.2 & 3.8 & 45.6 & -4.9 & 44.2 \\ 
  BOOT & -2.2 & 16.1 & 18.9 & 27.4 & -7.9 & 31.2 & -27.6 & 34.6 \\   
   \bottomrule
\end{tabular*}}
}
\end{table}

\newpage
\begin{table}[H]
{
\caption{ Coverage error rates of $95$\% level asymptotic bootstrap confidence intervals under Poisson sampling.}
\centering
\footnotesize
\setlength\tabcolsep{2pt}
\resizebox{\textwidth}{!}{
\begin{tabular*}{\textwidth}{@{\extracolsep{\fill}}*{14}{c}}
\toprule
  & \multicolumn{6}{c}{$F_0^{\mathrm{sym}}$} & \multicolumn{6}{c}{$F_0^{\mathrm{asym}}$}\\
  \cmidrule(l){2-7} \cmidrule(l){8-13}
   &\multicolumn{3}{c}{${\xi}_{0.50}$} & \multicolumn{3}{c}{${\xi}_{0.75}$} & \multicolumn{3}{c}{${\xi}_{0.50}$} & \multicolumn{3}{c}{${\xi}_{0.75}$}\\
    \cmidrule(l){2-4} \cmidrule(l){5-7} \cmidrule(l){8-10} \cmidrule(l){11-13}
 & $\text{L\%}$ & $\text{U\%}$ & $\text{L+U\%}$ & $\text{L\%}$ & $\text{U\%}$ & $\text{L+U\%}$ & $\text{L\%}$ & $\text{U\%}$ & $\text{L+U\%}$ & $\text{L\%}$ & $\text{U\%}$ & $\text{L+U\%}$ \\ 
\midrule
 \multicolumn{13}{l}{ (i) $n=100$, $f=7$\%, $N = 1~428 $}\\
\addlinespace
UNSMTHD & 2.1 & 3.1 & 5.1 & 1.9 & 3.2 & 5.1 & 3.7 & 2.7 & 6.4 & 2.2 & 3.2 & 5.5 \\  
BOOT & 1.0 & 1.5 & 2.5 & 1.3 & 3.1 & 4.4 & 2.0 & 1.2 & 3.2 & 2.9 & 1.5 & 4.3 \\ 
 \addlinespace
   \multicolumn{13}{l}{(ii) $n=100$, $f=30$\%, $N = 333$}\\
   \addlinespace
    UNSMTHD & 1.7 & 2.8 & 4.5 & 3.2 & 3.4 & 6.6 & 2.5 & 3.1 & 5.6 & 1.7 & 10.0 & 11.7 \\ 
  BOOT & 0.5 & 2.8 & 3.3 & 1.1 & 4.6 & 5.7 & 0.9 & 3.4 & 4.3 & 1.3 & 8.6 & 9.8 \\ 
  \addlinespace
   \multicolumn{13}{l}{(iii) $n=500$, $f=7$\%, $N = 7~142$}\\
   \addlinespace
UNSMTHD & 2.7 & 2.5 & 5.2 & 2.5 & 2.9 & 5.3 & 2.6 & 2.3 & 5.0 & 2.2 & 3.5 & 5.8 \\  
 BOOT & 1.7 & 2.3 & 4.0 & 1.4 & 2.4 & 3.8 & 2.6 & 2.6 & 5.2 & 2.6 & 3.1 & 5.7 \\ 
  \addlinespace
\multicolumn{13}{l}{(iv) $n=500$, $f=30$\%, $N = 1~666$}\\
\addlinespace
    UNSMTHD & 1.9 & 3.9 & 5.9 & 1.6 & 4.1 & 5.7 & 2.8 & 4.5 & 7.2 & 1.1 & 8.3 & 9.4 \\ 
  BOOT & 1.5 & 5.0 & 6.5 & 1.2 & 4.2 & 5.4 & 3.3 & 4.0 & 7.3 & 1.2 & 7.5 & 8.8 \\
   \bottomrule
\end{tabular*}}
\begin{tablenotes}
\footnotesize
    \item The 95\% confidence intervals for the two-tail 5.0\% and the one-tail 2.5\% nominal error rates are [4.0; 6.0]\% and [1.8; 3.2]\%, respectively.
  \end{tablenotes}
\label{tab:asymptotic_poisson}
}
\end{table}

\begin{table}[H]
{
\caption{Coverage error rates of $95$\% level basic bootstrap confidence intervals under Poisson sampling.}
\centering
\footnotesize
\footnotesize
\setlength\tabcolsep{2pt}
\begin{threeparttable}
\begin{tabular*}{\textwidth}{@{\extracolsep{\fill}}*{14}{c}}
\toprule
  & \multicolumn{6}{c}{$F_0^{\mathrm{sym}}$} & \multicolumn{6}{c}{$F_0^{\mathrm{asym}}$}\\
  \cline{2-7} \cline{8-13}
   &\multicolumn{3}{c}{${\xi}_{0.50}$} & \multicolumn{3}{c}{${\xi}_{0.75}$} & \multicolumn{3}{c}{${\xi}_{0.50}$} & \multicolumn{3}{c}{${\xi}_{0.75}$}\\
    \cline{2-4} \cline{5-7} \cline{8-10} \cline{11-13}
 & $\text{L\%}$ & $\text{U\%}$ & $\text{L+U\%}$ & $\text{L\%}$ & $\text{U\%}$ & $\text{L+U\%}$ & $\text{L\%}$ & $\text{U\%}$ & $\text{L+U\%}$ & $\text{L\%}$ & $\text{U\%}$ & $\text{L+U\%}$ \\ 
\midrule
 \multicolumn{13}{l}{ (i) $n=100$, $f=7$\%, $N = 1~428 $}\\
 \addlinespace
UNSMTHD &  4.2 & 9.4 & 13.6 & 4.2 & 6.2 & 10.5 & 7.1 & 11.5 & 18.6 & 7.7 & 12.3 & 20.0 \\ 
 BOOT & 1.1 & 1.6 & 2.7 & 1.5 & 3.4 & 4.8 & 1.6 & 3.0 & 4.6 & 2.2 & 2.2 & 4.5 \\   
   \addlinespace
   \multicolumn{13}{l}{(ii) $n=100$, $f=30$\%, $N = 333$}\\
    \addlinespace
  UNSMTHD & 3.5 & 6.1 & 9.6 & 7.6 & 8.8 & 16.4 & 5.4 & 9.3 & 14.8 & 5.9 & 28.9 & 34.8 \\  
  BOOT & 0.5 & 2.9 & 3.5 & 1.2 & 4.7 & 5.9 & 0.7 & 4.5 & 5.2 & 0.9 & 10.8 & 11.8 \\  
   \addlinespace
   \multicolumn{13}{l}{(iii) $n=500$, $f=7$\%, $N = 7~142$}\\
    \addlinespace
 UNSMTHD & 5.0 & 3.6 & 8.6 & 4.2 & 3.7 & 8.0 & 5.0 & 6.6 & 11.5 & 3.7 & 7.0 & 10.8 \\ 
  BOOT & 1.8 & 2.4 & 4.2 & 1.7 & 2.4 & 4.0 & 2.1 & 3.9 & 5.9 & 2.2 & 3.8 & 6.0 \\ 
  \addlinespace
\multicolumn{13}{l}{(iv) $n=500$, $f=30$\%, $N = 1~666$}\\
 \addlinespace
 UNSMTHD & 2.6 & 7.1 & 9.8 & 3.3 & 6.1 & 9.4 & 5.3 & 6.9 & 12.2 & 4.3 & 13.2 & 17.5 \\
  BOOT & 1.6 & 5.0 & 6.6 & 1.3 & 4.1 & 5.4 & 3.0 & 4.4 & 7.4 & 1.1 & 7.4 & 8.6 \\ 
   \hline
\end{tabular*}
\begin{tablenotes}
\footnotesize
    \item The 95\% confidence intervals for the two-tail 5.0\% and the one-tail 2.5\% nominal error rates are [4.0; 6.0]\% and [1.8; 3.2]\%, respectively.
  \end{tablenotes}
\end{threeparttable}
\label{tab:basic_poisson}
}
\end{table}

\newpage
\section{{Results under Randomized Systematic Proportional-to-Size Sampling}}
\label{appendix:syspps}

{In this subsection, we evaluate the performance of the UNSMTHD and BOOT methods under randomized systematic proportional-to-size sampling, implemented using the R code from \cite{wu2020sampling}. Table \ref{tab:rrmse_syspps} summarizes bias and instability metrics for the MSE estimators across the scenarios described in Subsection \ref{section:generation}. Similar to Poisson sampling, smoothing substantially improves the stability of the bootstrap-based mean squared error estimator across almost all scenarios. However, for the median and the $F_0^{\mathrm{asym}}$, the smoothed bootstrap estimator exhibits larger instability than the unsmoothed counterpart in smaller sample sizes, and the difference is more noticeable when the sample is drawn from a smaller population ($f=30$\%).}

{Tables \ref{tab:asymptotic_syspps} and \ref{tab:basic_syspps} report performance metrics for the asymptotic and basic bootstrap confidence intervals, respectively. Smoothed asymptotic intervals tend to be overly conservative due to their increased width, yielding little benefit over their unsmoothed counterparts, which already achieve near-nominal two-tail error rates in most cases. In contrast, smoothing has a pronounced impact on basic intervals: the unsmoothed version often leads to severe undercoverage, while the increased length of the smoothed intervals helps mitigate this issue. The lackluster one-tail error performance previously noted under SRSWOR and Poisson sampling is also evident here.}

\begin{table}[H]
{
\caption{\small Bias and instability of mean squared error estimators under randomized systematic proportional-to-size sampling.}
\label{tab:rrmse_syspps}
\centering
\footnotesize
\setlength\tabcolsep{2pt}
\resizebox{\textwidth}{!}{%
\begin{tabular*}{\textwidth}{@{\extracolsep{\fill}}*{9}{c}}
  \toprule
  & \multicolumn{4}{c}{$F_0^{\mathrm{sym}}$} & \multicolumn{4}{c}{$F_0^{\mathrm{asym}}$}\\
  \cmidrule(l){2-5} \cmidrule(l){6-9}
   &\multicolumn{2}{c}{$\mathrm{Var}_p(\hat{\xi}_{0.50})$} & \multicolumn{2}{c}{$\mathrm{Var}_p(\hat{\xi}_{0.75})$} & \multicolumn{2}{c}{$\mathrm{Var}_p(\hat{\xi}_{0.50})$} & \multicolumn{2}{c}{$\mathrm{Var}_p(\hat{\xi}_{0.75})$}\\
    \cmidrule(l){2-3} \cmidrule(l){4-5} \cmidrule(l){6-7} \cmidrule(l){8-9}
Method& Bias\% & RRMSE\% & Bias\% & RRMSE\% & Bias\% & RRMSE\% & Bias\% & RRMSE\% \\ 
  \midrule
 \multicolumn{4}{l}{ (i) $n=100$, $f=7$\%, $N = 1~428 $}\\
 \addlinespace
UNSMTHD & 11.5 & 51.1 & 20.9 & 61.5 & 15.1 & 73.9 & 18.0 & 95.9 \\ 
BOOT & 24.8 & 39.6 & 27.3 & 42.2 & 27.3 & 78.9 & 18.4 & 91.3 \\  
 \addlinespace
  \multicolumn{4}{l}{(ii) $n=100$, $f=30$\%, $N = 333$}\\
   \addlinespace
  UNSMTHD  & 21.1 & 59.2 & 15.2 & 57.5 & 7.8 & 69.7 & 6.4 & 67.6 \\ 
  BOOT & 35.4 & 46.1 & 5.3 & 23.0 & 27.9 & 82.2 & -10.0 & 54.8 \\ 
   \addlinespace
 \multicolumn{4}{l}{(iv) $n=500$, $f=30$\%, $N = 1~666$}\\
  \addlinespace
  UNSMTHD& 1.5 & 32.4 & 7.5 & 35.1 & 2.5 & 38.8 & 1.9 & 43.9 \\ 
  BOOT  & -4.9 & 16.1 & 17.2 & 25.9 & -5.8 & 23.8 & -22.6 & 31.5 \\    
   \bottomrule
\end{tabular*}}
}

\end{table}

\newpage
\begin{table}[H]
{
\caption{ Coverage error rates of $95$\% level asymptotic bootstrap confidence intervals under randomized systematic proportional-to-size sampling.}
\centering
\footnotesize
\setlength\tabcolsep{2pt}
\resizebox{\textwidth}{!}{
\begin{tabular*}{\textwidth}{@{\extracolsep{\fill}}*{14}{c}}
\toprule
  & \multicolumn{6}{c}{$F_0^{\mathrm{sym}}$} & \multicolumn{6}{c}{$F_0^{\mathrm{asym}}$}\\
  \cmidrule(l){2-7} \cmidrule(l){8-13}
   &\multicolumn{3}{c}{${\xi}_{0.50}$} & \multicolumn{3}{c}{${\xi}_{0.75}$} & \multicolumn{3}{c}{${\xi}_{0.50}$} & \multicolumn{3}{c}{${\xi}_{0.75}$}\\
    \cmidrule(l){2-4} \cmidrule(l){5-7} \cmidrule(l){8-10} \cmidrule(l){11-13}
 & $\text{L\%}$ & $\text{U\%}$ & $\text{L+U\%}$ & $\text{L\%}$ & $\text{U\%}$ & $\text{L+U\%}$ & $\text{L\%}$ & $\text{U\%}$ & $\text{L+U\%}$ & $\text{L\%}$ & $\text{U\%}$ & $\text{L+U\%}$ \\ 
\midrule
 \multicolumn{13}{l}{ (i) $n=100$, $f=7$\%, $N = 1~428 $}\\
\addlinespace
UNSMTHD & 1.7 & 4.2 & 5.9 & 1.8 & 3.3 & 5.1 & 3.8 & 2.9 & 6.7 & 2.9 & 3.8 & 6.6 \\ 
BOOT & 1.1 & 1.9 & 3.0 & 1.2 & 2.8 & 4.0 & 2.0 & 1.9 & 3.9 & 2.5 & 2.0 & 4.5 \\ 
 \addlinespace
   \multicolumn{13}{l}{(ii) $n=100$, $f=30$\%, $N = 333$}\\
   \addlinespace
    UNSMTHD & 2.0 & 2.6 & 4.7 & 2.5 & 3.1 & 5.7 & 3.4 & 3.4 & 6.7 & 2.0 & 7.0 & 8.9 \\ 
  BOOT & 0.8 & 2.9 & 3.6 & 1.0 & 4.6 & 5.6 & 0.9 & 3.9 & 4.8 & 1.4 & 7.4 & 8.8 \\ 
  \addlinespace
\multicolumn{13}{l}{(iv) $n=500$, $f=30$\%, $N = 1~666$}\\
\addlinespace
    UNSMTHD & 1.8 & 3.6 & 5.5 & 2.1 & 3.8 & 5.9 & 3.2 & 2.5 & 5.7 & 2.2 & 5.6 & 7.8 \\ 
  BOOT & 1.4 & 4.9 & 6.3 & 1.1 & 3.7 & 4.8 & 3.2 & 2.3 & 5.5 & 2.1 & 5.5 & 7.5 \\ 
   \bottomrule
\end{tabular*}}
\begin{tablenotes}
\footnotesize
    \item The 95\% confidence intervals for the two-tail 5.0\% and the one-tail 2.5\% nominal error rates are [4.0; 6.0]\% and [1.8; 3.2]\%, respectively.
  \end{tablenotes}
\label{tab:asymptotic_syspps}
}
\end{table}

\begin{table}[H]
{
\caption{Coverage error rates of $95$\% level basic bootstrap confidence intervals under randomized systematic proportional-to-size sampling.}
\centering
\footnotesize
\footnotesize
\setlength\tabcolsep{2pt}
\begin{threeparttable}
\begin{tabular*}{\textwidth}{@{\extracolsep{\fill}}*{14}{c}}
\toprule
  & \multicolumn{6}{c}{$F_0^{\mathrm{sym}}$} & \multicolumn{6}{c}{$F_0^{\mathrm{asym}}$}\\
  \cline{2-7} \cline{8-13}
   &\multicolumn{3}{c}{${\xi}_{0.50}$} & \multicolumn{3}{c}{${\xi}_{0.75}$} & \multicolumn{3}{c}{${\xi}_{0.50}$} & \multicolumn{3}{c}{${\xi}_{0.75}$}\\
    \cline{2-4} \cline{5-7} \cline{8-10} \cline{11-13}
 & $\text{L\%}$ & $\text{U\%}$ & $\text{L+U\%}$ & $\text{L\%}$ & $\text{U\%}$ & $\text{L+U\%}$ & $\text{L\%}$ & $\text{U\%}$ & $\text{L+U\%}$ & $\text{L\%}$ & $\text{U\%}$ & $\text{L+U\%}$ \\ 
\midrule
 \multicolumn{13}{l}{ (i) $n=100$, $f=7$\%, $N = 1~428 $}\\
 \addlinespace
UNSMTHD  & 4.5 & 9.4 & 13.9 & 4.2 & 7.0 & 11.2 & 7.8 & 12.6 & 20.3 & 7.4 & 12.5 & 19.9 \\ 
 BOOT& 1.2 & 1.9 & 3.1 & 1.3 & 2.9 & 4.2 & 1.8 & 2.9 & 4.7 & 1.9 & 2.6 & 4.5 \\  
   \addlinespace
   \multicolumn{13}{l}{(ii) $n=100$, $f=30$\%, $N = 333$}\\
    \addlinespace
  UNSMTHD & 3.9 & 5.8 & 9.7 & 7.5 & 8.6 & 16.2 & 5.6 & 8.9 & 14.5 & 7.4 & 23.6 & 31.1 \\ 
  BOOT & 0.8 & 2.6 & 3.4 & 1.2 & 4.3 & 5.6 & 0.5 & 5.3 & 5.9 & 0.9 & 9.8 & 10.7 \\ 
  \addlinespace
\multicolumn{13}{l}{(iv) $n=500$, $f=30$\%, $N = 1~666$}\\
 \addlinespace
 UNSMTHD & 2.2 & 6.6 & 8.8 & 3.2 & 5.1 & 8.3 & 5.6 & 4.9 & 10.5 & 5.9 & 11.6 & 17.5 \\ 
  BOOT & 1.5 & 5.0 & 6.5 & 1.1 & 3.7 & 4.8 & 2.5 & 2.8 & 5.3 & 1.5 & 6.1 & 7.6 \\ 
   \hline
\end{tabular*}
\begin{tablenotes}
\footnotesize
    \item The 95\% confidence intervals for the two-tail 5.0\% and the one-tail 2.5\% nominal error rates are [4.0; 6.0]\% and [1.8; 3.2]\%, respectively.
  \end{tablenotes}
\end{threeparttable}
\label{tab:basic_syspps}
}
\end{table}

\newpage
\section{{Additional results for SRSWOR}}
\label{appendix:srswor}
For the SRSWOR design, we considered two additional scenarios for the underlying distribution $F_0$: $\mathcal{N}(0,1)$ and $\mathrm{Lognormal}(0,1)$. See Table \ref{tab:grille_eassr2} for the optimal constants along with the constant estimated by the plug-in method for two superpopulations $F_0$. Note that in the case of the median of the $\mathrm{Lognormal}(0,1)$ distribution, the minimizer of the asymptotic mean squared error of the bootstrap variance estimator, based on a second-order Taylor expansion, does not exist given that $f_0''(\xi_{0.50}) - f_0'(\xi_{0.50})^2f_0(\xi_{0.50})^{-1}=0$. The same grid as the one used for $F_0=\mathcal{N}(0,1)$ was selected for this scenario after verifying by simulation that a minimum mean squared error was achieved empirically for $\hat{V}^*_h$ over the range of fixed values in $\mathcal{H}(n)$. 

Figures \ref{fig:instability_norm2} and \ref{fig:instability_lnorm2} show the range considered for the grids $\mathcal{H}(n)$ for the different quantiles and superpopulations considered under SRSWOR. The bias and relative instability obtained over the 2,000 simulations are reported in Table \ref{tab:rrmse_eassr2} for each of the bandwidth selection approaches. As well, RRMSE\% values for all three methods considered are compared to a relative instability curve obtained for a fixed bandwidth grid in Figure \ref{fig:instability_norm2} for the $\mathcal{N}(0,1)$ superpopulation and in Figure \ref{fig:instability_lnorm2} for the $\mathrm{Lognormal}(0,1)$ superpopulation. Results show that, compared with the UNSMTHD method, substantial reductions in RRMSE\% can be achieved from the smoothed pseudo-population bootstrap approach with a data-driven selection of the bandwidth.

\begin{table}[H]
\centering
\caption{Optimal constants used to construct the bandwidth grids $\mathcal{H}(n)$ in additional simulations}
\begin{threeparttable}
\begin{tabular}{l c c }
\hline
Scenario  & $C_{\text{opt}}^{\phi}(\xi_p)$ & $C_{\text{norm}}^{\phi}(\xi_p)$ \\
\hline
${\xi}_{0.50}, \ \mathcal{N}(0,1)$ & 0.93 & 0.93
\\
${\xi}_{0.75},\ \mathcal{N}(0,1)$ & 0.98 & 0.98 
\\
${\xi}_{0.50},\ \mathrm{Lognormal}(0,1)$  & $\infty$ &2.03 
\\
${\xi}_{0.75},\ \mathrm{Lognormal}(0,1)$  &2.24  &2.02 
\\
\hline
\end{tabular}
\end{threeparttable}
\label{tab:grille_eassr2}
\end{table}

\begin{table}[H]
\caption{\small Bias and instability of variance estimators under simple random sampling without replacement.}
\centering
\footnotesize
\setlength\tabcolsep{2pt}
\resizebox{\textwidth}{!}{%
\begin{tabular*}{\textwidth}{@{\extracolsep{\fill}}*{9}{c}}
  \toprule
  & \multicolumn{4}{c}{$F_0 = \mathcal{N}(0,1)$} & \multicolumn{4}{c}{$F_0 = \mathrm{Lognormal}(0,1)$}\\
  \cmidrule(l){2-5} \cmidrule(l){6-9}
   &\multicolumn{2}{c}{$\mathrm{Var}_p(\hat{\xi}_{0.50})$} & \multicolumn{2}{c}{$\mathrm{Var}_p(\hat{\xi}_{0.75})$} & \multicolumn{2}{c}{$\mathrm{Var}_p(\hat{\xi}_{0.50})$} & \multicolumn{2}{c}{$\mathrm{Var}_p(\hat{\xi}_{0.75})$}\\
    \cmidrule(l){2-3} \cmidrule(l){4-5} \cmidrule(l){6-7} \cmidrule(l){8-9}
Method& Bias\% & RRMSE\% & Bias\% & RRMSE\% & Bias\% & RRMSE\% & Bias\% & RRMSE\% \\ 
  \midrule
 \multicolumn{4}{l}{ (i) $n=100$, $f=7$\%, $N = 1~428 $}\\
 \addlinespace
UNSMTHD & 16.7 & 55.7 & 25.2 & 68.3 & 24.3 & 69.9 & 31.9 & 86.1 \\ 
BOOT & 25.7 & 39.5 & 24.8 & 41.4 & 42.7 & 66.6 & 32.3 & 54.3 \\ 
PLUG-IN & 17.1 & 30.9 & 18.9 & 32.8 & 114.3 & 146.2 & 18.5 & 47.4 \\ 
 \addlinespace
  \multicolumn{4}{l}{(ii) $n=100$, $f=30$\%, $N = 333$}\\
   \addlinespace
  UNSMTHD & 14.7 & 51.6 & 24.5 & 65.3 & 28.0 & 74.6 & 33.9 & 89.3 \\ 
  BOOT & 37.3 & 51.1 & 28.9 & 39.2 & 79.8 & 105.2 & 28.7 & 46.5 \\ 
  PLUG-IN & 31.6 & 41.6 & 23.4 & 34.5 & 161.0 & 174.4 & 17.9 & 43.9 \\ 
   \addlinespace
 \multicolumn{4}{l}{(iii) $n=500$, $f=7$\%, $N = 7~142$}\\
  \addlinespace
 UNSMTHD & 4.4 & 32.3 & 0.7 & 34.3 & 2.6 & 33.5 & 7.7 & 41.5 \\ 
 BOOT & 6.9 & 18.0 & -0.2 & 16.1 & -3.7 & 17.4 & 1.0 & 17.8 \\ 
  PLUG-IN & 6.7 & 14.9 & -0.3 & 13.2 & 20.6 & 27.3 & -1.1 & 18.9 \\ 
   \addlinespace
 \multicolumn{4}{l}{(iv) $n=500$, $f=30$\%, $N = 1~666$}\\
  \addlinespace
  UNSMTHD & 1.3 & 31.5 & 8.9 & 35.1 & 19.0 & 42.2 & 12.3 & 39.8 \\ 
  BOOT & 1.6 & 15.3 & 13.3 & 20.2 & 33.3 & 43.4 & 26.1 & 33.4 \\ 
  PLUG-IN & 2.5 & 12.4 & 12.7 & 18.9 & 96.1 & 99.9 & 24.9 & 31.5 \\ 
   \bottomrule
\end{tabular*}}
\label{tab:rrmse_eassr2}
\end{table}

\begin{figure}
\centering
\small{Instability curves for $\hat{V}_{\hat{h}}\left(\hat{\xi}_{0.50}\right)$ with underlying distribution $\mathcal{N}(0,1)$}\\
\includegraphics[scale=0.55]{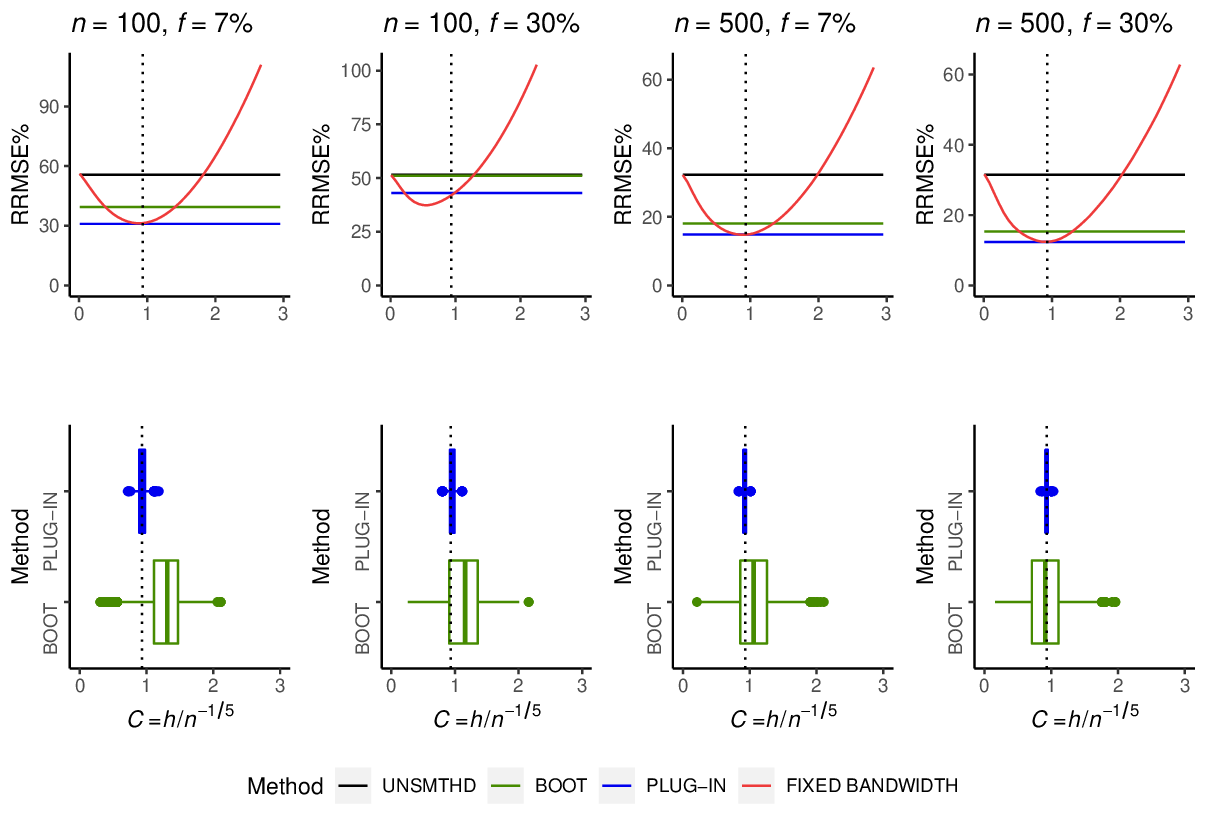}\\
\small{Instability curves for $\hat{V}_{\hat{h}}\left(\hat{\xi}_{0.75}\right)$ with underlying distribution $\mathcal{N}(0,1)$}\\
\includegraphics[scale=0.55]{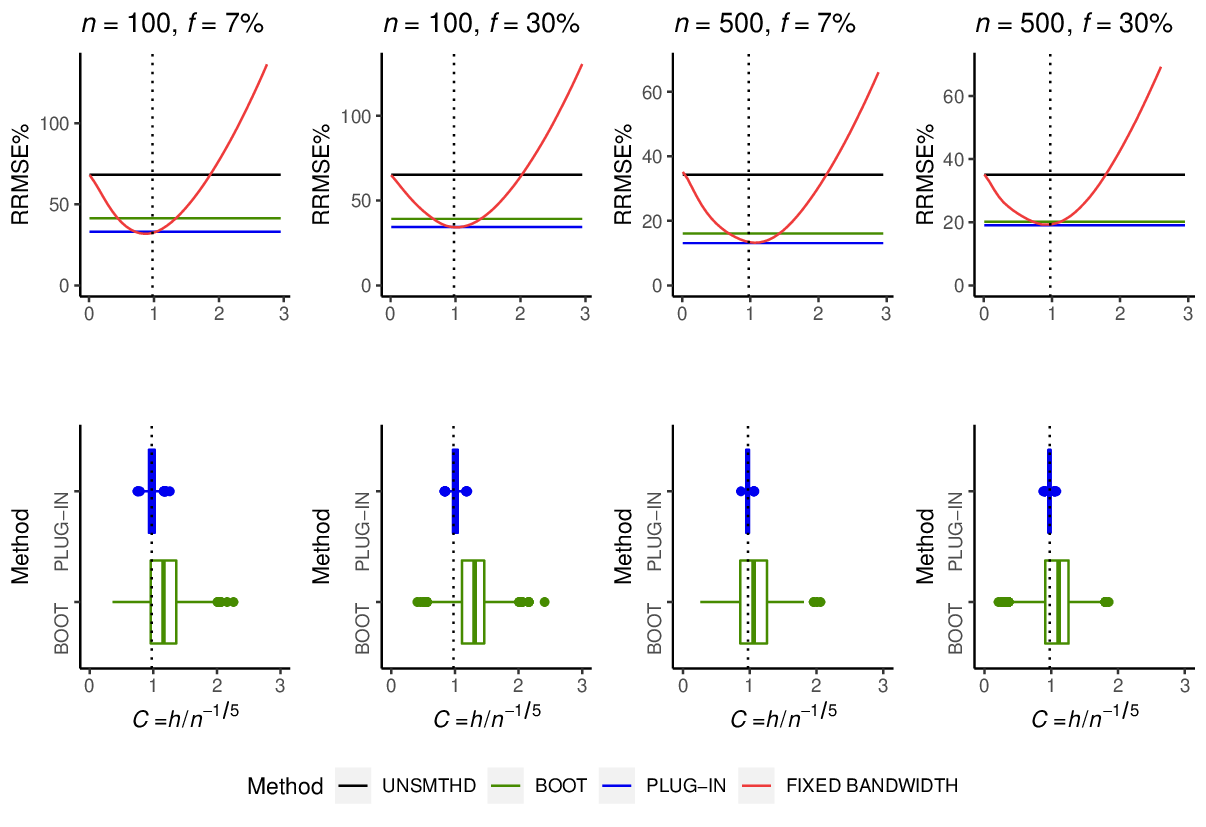}\\
\caption{ Instability (RRMSE\%) curves of bootstrap mean squared error estimators for four approaches and boxplots of selected bandwidths $\hat{C} = \hat{h}/n^{-1/5}$ for the BOOT and PLUG-IN approaches across $R = $ 2,000 simulated SRSWOR samples from a single finite population with ($n$, $f$) = (100, 0.07), (100, 0.30), (500, 0.07), (500, 0.30) and underlying distribution $\mathcal{N}(0,1)$. The red curve in the plots corresponds to the RRMSE\% of $\hat{V}_h$, where $h = C \cdot n^{-1/5}$ and $C$ is a known (fixed) constant. The solid black line is the RRMSE\% of the unsmoothed (standard) bootstrap method ($\hat{h} = 0$). The solid blue and green lines pertain to the PLUG-IN and BOOT bandwidth selection methods, respectively. The vertical dotted line corresponds to the value of the optimal bandwidth constant given in  (\ref{eqn:copt}) for the standard normal distribution. }

\label{fig:instability_norm2}
\end{figure}

\begin{figure}
\centering
\small{Instability curves for $\hat{V}_{\hat{h}}\left(\hat{\xi}_{0.50}\right)$ with underlying distribution $\mathrm{Lognormal}(0,1)$}\\
\includegraphics[scale=0.6]{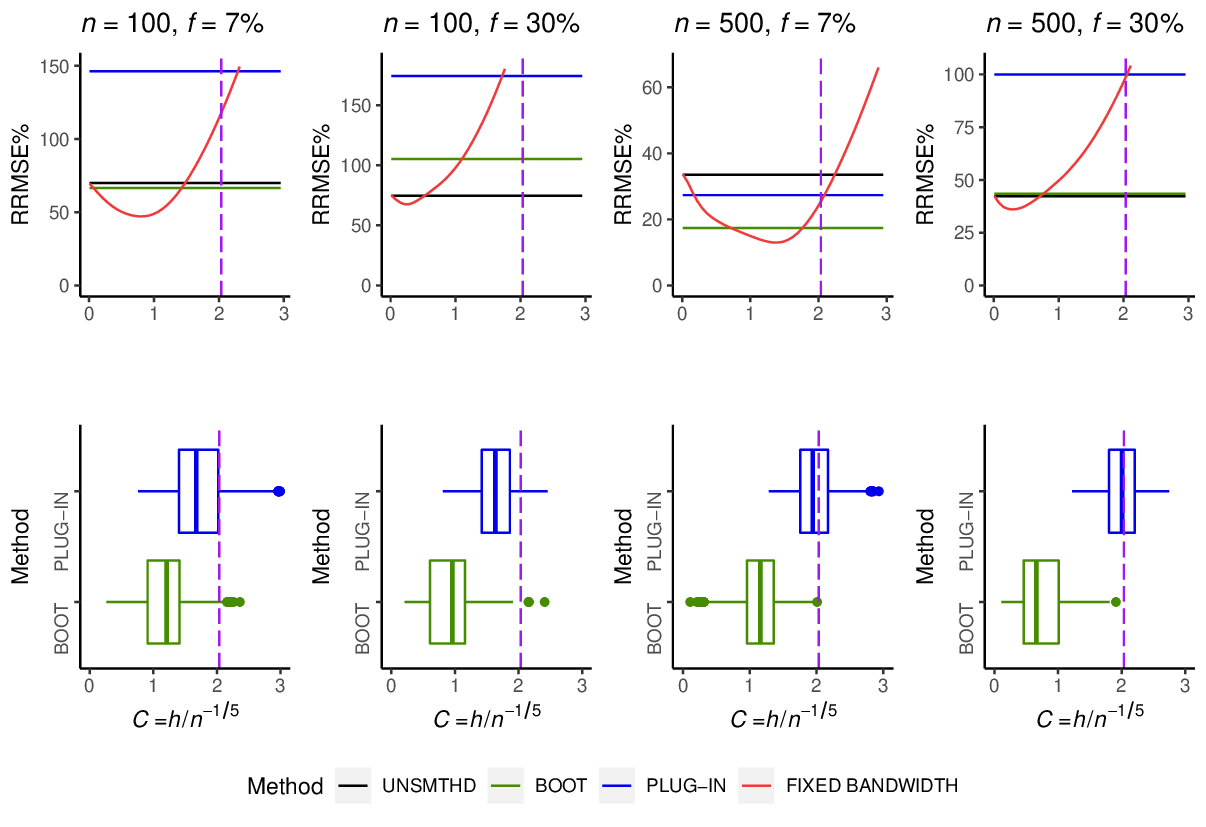}\\
\small{Instability curves for $\hat{V}_{\hat{h}}\left(\hat{\xi}_{0.75}\right)$ with underlying distribution $\mathrm{Lognormal}(0,1)$}\\
\includegraphics[scale=0.6]{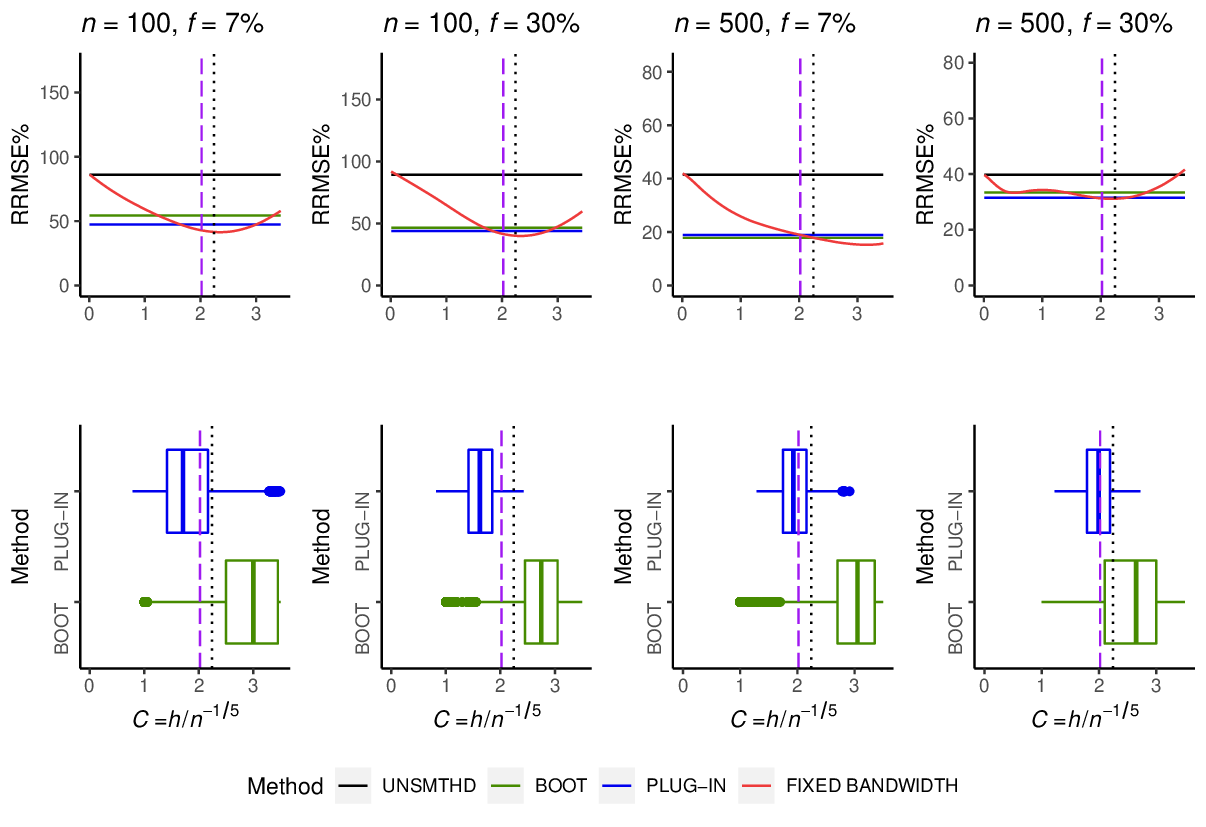}\\
\caption{ Instability (RRMSE\%) curves of bootstrap mean squared error estimators for four approaches and boxplots of selected bandwidths $\hat{C} = \hat{h}/n^{-1/5}$ for the BOOT and PLUG-IN approaches across $R = $ 2,000 simulated SRSWOR samples from a single finite population with ($n$, $f$) = (100, 0.07), (100, 0.30), (500, 0.07), (500, 0.30) and underlying distribution $\mathrm{Lognormal}(0,1)$. The red curve in the plots corresponds to the RRMSE\% of $\hat{V}_h$, where $h = C \cdot n^{-1/5}$ and $C$ is a known (fixed) constant. The solid black line is the RRMSE\% of the unsmoothed (standard) bootstrap variance estimator ($\hat{h} = 0$). The solid blue and green lines pertain to the PLUG-IN and BOOT bandwidth selection methods, respectively. The vertical dotted black line corresponds to the value of the optimal bandwidth constant given in  (\ref{eqn:copt}) for the distribution $\mathrm{Lognormal}(0,1)$ (note that it does not exist in the case of the median). The vertical dashed purple line is the optimal bandwidth constant for normally distributed observations given in (\ref{eqn:coptnorm}), which always exists and is estimated in the PLUG-IN method.  }
\label{fig:instability_lnorm2}
\end{figure}

\begin{table}[H]
\caption{ Coverage error rates of $95$\% level asymptotic bootstrap confidence intervals under simple random sampling without replacement.}
\label{tab:asymptotic_srs2}
\centering
\footnotesize
\begin{threeparttable}
\setlength\tabcolsep{2pt}
\resizebox{\textwidth}{!}{
\begin{tabular*}{\textwidth}{@{\extracolsep{\fill}}*{14}{c}}
\toprule
  & \multicolumn{6}{c}{$F_0 = \mathcal{N}(0,1)$} & \multicolumn{6}{c}{$F_0 = \mathrm{Lognormal}(0,1)$}\\
  \cmidrule(l){2-7} \cmidrule(l){8-13}
   &\multicolumn{3}{c}{${\xi}_{0.50}$} & \multicolumn{3}{c}{${\xi}_{0.75}$} & \multicolumn{3}{c}{${\xi}_{0.50}$} & \multicolumn{3}{c}{${\xi}_{0.75}$}\\
    \cmidrule(l){2-4} \cmidrule(l){5-7} \cmidrule(l){8-10} \cmidrule(l){11-13}
 & $\text{L\%}$ & $\text{U\%}$ & $\text{L+U\%}$ & $\text{L\%}$ & $\text{U\%}$ & $\text{L+U\%}$ & $\text{L\%}$ & $\text{U\%}$ & $\text{L+U\%}$ & $\text{L\%}$ & $\text{U\%}$ & $\text{L+U\%}$ \\ 
\midrule
 \multicolumn{13}{l}{ (i) $n=100$, $f=7$\%, $N = 1~428 $}\\
\addlinespace
UNSMTHD & 2.4 & 2.7 & 5.1 & 1.8 & 3.1 & 5.0 & 1.4 & 3.8 & 5.3 & 0.8 & 4.6 & 5.3 \\ 
BOOT & 0.9 & 2.6 & 3.5 & 1.8 & 1.6 & 3.4 & 0.6 & 2.4 & 3.0 & 0.5 & 3.8 & 4.2 \\ 
 PLUG-IN & 1.0 & 3.0 & 4.0 & 1.9 & 1.9 & 3.8 & 0.2 & 0.4 & 0.7 & 0.6 & 5.3 & 5.9 \\ 
 \addlinespace
   \multicolumn{13}{l}{(ii) $n=100$, $f=30$\%, $N = 333$}\\
   \addlinespace
    UNSMTHD & 1.9 & 2.9 & 4.8 & 1.8 & 1.8 & 3.6 & 0.8 & 1.6 & 2.4 & 0.6 & 5.9 & 6.5 \\ 
  BOOT & 0.9 & 2.2 & 3.1 & 1.3 & 2.3 & 3.6 & 0.2 & 1.5 & 1.7 & 1.0 & 4.8 & 5.9 \\ 
  PLUG-IN & 0.8 & 1.9 &  2.8 & 1.6 & 2.5 & 4.1 & $<0.1$ & 0.4 & 0.4 & 0.9 & 6.3 & 7.2 \\  
  \addlinespace
   \multicolumn{13}{l}{(iii) $n=500$, $f=7$\%, $N = 7~142$}\\
   \addlinespace
UNSMTHD & 2.8 & 2.1 & 4.9 & 1.5 & 5.2 & 6.8 & 3.2 & 2.2 & 5.4 & 1.6 & 2.8 & 4.4 \\ 
 BOOT & 1.9 & 1.6 & 3.5 & 0.7 & 4.5 & 5.2 & 2.4 & 1.8 & 4.2 & 1.8 & 2.8 & 4.5 \\ 
  PLUG-IN & 2.0 & 1.6 & 3.5 & 0.7 & 4.3 & 5.0 & 1.3 & 0.9 & 2.2 & 1.6 & 3.1 & 4.8 \\ 
  \addlinespace
\multicolumn{13}{l}{(iv) $n=500$, $f=30$\%, $N = 1~666$}\\
\addlinespace
    UNSMTHD & 3.9 & 1.8 & 5.6 & 1.6 & 3.1 & 4.8 & 1.2 & 2.2 & 3.5 & 1.8 & 2.6 & 4.3 \\ 
  BOOT & 1.8 & 1.5 & 3.4 & 1.0 & 2.8 & 3.8 & 0.9 & 2.8 & 3.8 & 0.7 & 2.4 & 3.0 \\ 
  PLUG-IN & 1.5 & 1.5 & 3.0 & 0.9 & 2.7 & 3.6 & 0.2 & 0.9 & 1.1 & 0.5 & 2.5 & 3.0 \\ 
   \bottomrule
\end{tabular*}}
\begin{tablenotes}
\footnotesize
    \item The 95\% confidence intervals for the two-tail 5.0\% and the one-tail 2.5\% nominal error rates are [4.0; 6.0]\% and [1.8; 3.2]\%, respectively.
  \end{tablenotes}
\end{threeparttable}
\end{table}

\begin{table}[H]
\caption{Coverage error rates of $95$\% level basic bootstrap confidence intervals under simple random sampling without replacement.}
\centering
\footnotesize
\begin{threeparttable}
\begin{tabular*}{\textwidth}{@{\extracolsep{\fill}}*{14}{c}}
\toprule
  & \multicolumn{6}{c}{$F_0 = \mathcal{N}(0,1)$} & \multicolumn{6}{c}{$F_0 = \mathrm{Lognormal}(0,1)$}\\
  \cline{2-7} \cline{8-13}
   &\multicolumn{3}{c}{${\xi}_{0.50}$} & \multicolumn{3}{c}{${\xi}_{0.75}$} & \multicolumn{3}{c}{${\xi}_{0.50}$} & \multicolumn{3}{c}{${\xi}_{0.75}$}\\
    \cline{2-4} \cline{5-7} \cline{8-10} \cline{11-13}
 & $\text{L\%}$ & $\text{U\%}$ & $\text{L+U\%}$ & $\text{L\%}$ & $\text{U\%}$ & $\text{L+U\%}$ & $\text{L\%}$ & $\text{U\%}$ & $\text{L+U\%}$ & $\text{L\%}$ & $\text{U\%}$ & $\text{L+U\%}$ \\ 
\midrule
 \multicolumn{13}{l}{ (i) $n=100$, $f=7$\%, $N = 1~428 $}\\
 \addlinespace
UNSMTHD & 4.5 & 6.3 & 10.8 & 6.0 & 6.2 & 12.2 & 2.5& 9.6 & 12.2 & 2.2 & 10.8 & 13.1 \\ 
 BOOT & 1.1 & 2.8 & 3.9 & 2.4 & 1.6 & 4.0 & 0.6 & 3.6 & 4.3 & 0.5 & 4.6 & 5.1 \\ 
  PLUG-IN & 1.4 & 3.2 & 4.6 & 2.6 & 1.9 & 4.5 & 0.2 & 1.0 & 1.2 & 0.4 & 7.3 & 7.7 \\ 
   \addlinespace
   \multicolumn{13}{l}{(ii) $n=100$, $f=30$\%, $N = 333$}\\
    \addlinespace
  UNSMTHD & 3.2 & 5.1 & 8.3 & 4.3 & 2.8 & 7.1 & 1.0 & 2.7 & 3.8 & 1.4 & 8.5 & 9.8 \\ 
  BOOT & 1.0 & 2.2 & 3.2 & 2.2 & 2.2 & 4.4 & $<0.1$ & 1.8 & 1.9 & 0.9 & 5.2 & 6.2 \\ 
  PLUG-IN & 0.9 & 2.1 & 3.0 & 2.4 & 2.2 & 4.7 & $<0.1$ &0.4 & 0.5 & 0.8 & 6.9 & 7.8 \\   
   \addlinespace
   \multicolumn{13}{l}{(iii) $n=500$, $f=7$\%, $N = 7~142$}\\
    \addlinespace
 UNSMTHD & 5.3 & 3.2 & 8.6 & 2.4 & 8.0 & 10.4 & 5.2 & 3.9 & 9.1 & 3.1 & 5.4 & 8.6 \\ 
  BOOT & 2.1 & 1.7 & 3.8 & 0.7 & 4.2 & 4.9 & 2.2 & 2.6 & 4.8 & 1.6 & 3.2 & 4.8 \\ 
 PLUG-IN & 2.1 & 1.7 & 3.8 & 1.0 & 4.3 & 5.3 & 1.1 & 1.3 & 2.4 & 1.2 & 3.5 & 4.8 \\ 
  \addlinespace
\multicolumn{13}{l}{(iv) $n=500$, $f=30$\%, $N = 1~666$}\\
 \addlinespace
 UNSMTHD & 6.6 & 1.7 & 8.2 & 2.5 & 4.0 & 6.6 & 1.5 & 3.5 & 5.0 & 2.1 & 2.8 & 5.0 \\ 
  BOOT & 1.9 & 1.6 & 3.6 & 1.1 & 2.8 & 3.9 & 0.8 & 3.2 & 4.0 & 0.5 & 2.5 & 3.0 \\ 
 PLUG-IN & 1.5 & 1.5 & 3.0 & 1.2 & 2.8 & 4.0 & 0.2 & 1.1 & 1.3 & 0.5 & 2.9 & 3.4 \\
   \hline
\end{tabular*}
\begin{tablenotes}
\footnotesize
    \item The 95\% confidence intervals for the two-tail 5.0\% and the one-tail 2.5\% nominal error rates are [4.0; 6.0]\% and [1.8; 3.2]\%, respectively.
  \end{tablenotes}
\end{threeparttable}
\label{tab:basic_srs2}
\end{table}

\newpage 
\section{{Smoothed pseudo-population algorithm for variance estimation}}
\label{appendix:smoothedppbvariance}
{As mentioned before, the bootstrap estimator in (\ref{eqn:bootstrap_estimator}) is an estimator of the mean squared error of $\hat{\theta}$ if $\hat{\theta}$ is biased for $\theta$. So far, mean squared error has been the focus of this paper. In this section, we present a version of the smoothed pseudo-population bootstrap algorithm if variance estimation for a biased estimator is of interest. The method presented in Algorithm \ref{algo:smoothedppbvariance} is a smoothed version of the algorithm presented in \cite{chauvet2007bootstrap} and discussed in \cite{mashreghi2016survey}.} 

\begin{algorithm}[h]

    \caption{Smoothed UEQPS Pseudo-Population Bootstrap for Variance Estimation}
  \begin{algorithmic}[1]
   \vspace*{0.5cm}
    \STATE Form $U^f$, the fixed part of the pseudo-population, by replicating each pair $(y_i, \pi_i)$ a total of $\floor{\pi_i^{-1}}$ times, with $\floor{x}$ being the largest integer less or equal to $x$.
    \STATE Complete the pseudo-population by drawing $U^{*c}$ according to the original survey design with inclusion probability equal to $\pi_i^{-1} - \floor{\pi_i^{-1}}$ for unit $(y_i, \pi_i), \ i \in S$, leading to the pseudo-population $U^* = U^f \cup U^{c*} = \{(y^*_i, \pi^*_i)\}_{ i = 1, \ldots, N^* }$ with possibly random size $N^*$, where $(y^*_i, \pi^*_i)$ corresponds to one of the original pairs of value of the variable and first-order inclusion probability in $S$.
    
    \STATE To obtain a smoothed pseudo-population $U^*_h$, compute $y^*_{i,h} = y^*_i + h\varepsilon^*_i$, where $\varepsilon^*_i \stackrel{\text{i.i.d.}}{\sim} K$, $i = 1, \ldots, N^*$, and $h$ is the smoothing parameter.
      \STATE Using the original sampling design, generate a bootstrap sample $S^*_h $ from $U^*_h$, but with inclusion probability $\pi'_i$ for unit $i \in U^*, i = 1, \ldots, N^*$, as defined in Section \ref{sec3}.
      \STATE Compute the smoothed bootstrap estimator, given by $\hat{\theta}^*_h = \theta(S^*_h)$.
      \STATE For $b=1,\ldots, B$, with $B$ large enough, repeat the Steps 3 to 5 so as to obtain the following distribution of smoothed bootstrap estimates:
\[\quad (\hat{\theta}_{1,h}^*, \ldots, \hat{\theta}_{B,h}^*)'.\]
leading to $$ \hat{V}^*_{B, h} = \frac{1}{B-1} \sum_{b=1}^B \left(\hat{\theta}^*_{b,h} - \hat{\theta}^*_{., h} \right)^2, \quad \text{where } \hat{\theta}^*_{.,h} = \frac{1}{B}\sum_{b=1}^B \hat{\theta}^*_{b, h}. $$
\STATE For $a=1,\ldots, A,$ repeat Steps 2 to 7 to obtain $\hat{V}^*_{B,1,h}, \ldots, \hat{V}^*_{B,A,h},$ leading to
$$ \hat{V}_h^* = \frac{1}{A}\sum_{a=1}^A \hat{V}^*_{B,a,h}.$$
  \end{algorithmic}
  \label{algo:smoothedppbvariance}
\end{algorithm}

{This approach ensures that the resulting $\hat{V}^*_h$ is a proper approximation of the variance instead of the mean squared error in the case of a biased estimator and does not require the explicit computation of the bootstrap parameter $\theta^*_h$ at any step. Additionally, this formulation may be more appropriate for constructing normal-based confidence
intervals. However, it is more computationally expensive, especially if used in conjunction with the bootstrap bandwidth selection method described in Section \ref{sec4}. }

\end{document}